\documentclass[useAMS, usenatbib]{mn2e}
\usepackage{graphicx}
\usepackage{amsmath}
\usepackage{amsfonts}
\usepackage{perpage} %the perpage package
\usepackage[setpagesize=false]{hyperref} %Optional but really cool hyperlinks everywhere.
\title{Effects of Lens Motion and Uneven Magnification on Image Spectra}{} %A short title will destroy the document!!!!! Leave it blank: use long title on headers.
\MakePerPage{footnote} %the perpage package command

\author[Indranil Banik \& Hongsheng Zhao]{Indranil Banik$^{1}$\thanks{Email: ib45@st-andrews.ac.uk}, Hongsheng Zhao$^{1}$\\
$^{1}$Scottish Universities Physics Alliance, University of St Andrews, North Haugh, St Andrews, Fife, KY16 9SS, UK}

\begin{document}
\maketitle
%\tableofcontents

%\newpage

%\clearpage
%\nolink{\relax\blfootnote{$\star$ E-mail: ib45@st-andrews.ac.uk}}  %DO NOT PUT A BLANK LINE AFTER THIS, OR MOVE COMMAND EARLIER!!!!!!!!!!!!!!!!!!!!!!!!
\begin{abstract}

%The tangential velocity of the Bullet Cluster may provide an important test of $\Lambda CDM$. It might be measured using the Moving Cluster Effect (MCE), which relies on multiple images of the same object having different redshifts due to the gravitational potential of a moving lens being time-dependent. Here, we describe the Differential Magnification Effect (DME), another mechanism for causing such redshift differences. We show that it might well lead to effects comparable to the MCE. However, we find that if detailed spectral line profiles were available, then the DME and MCE could be distinguished.

Counter to intuition, the images of an extended galaxy lensed by a moving galaxy cluster should have slightly different spectra in any metric gravity theory. This is mainly for two reasons. One relies on the gravitational potential of a moving lens being time-dependent (the $\text{Moving}$ $\text{Cluster}$ $\text{Effect}$, $\text{MCE}$). The other is due to uneven magnification across the extended, rotating source (the $\text{Differential}$ $\text{Magnification}$ $\text{Effect}$, $\text{DME}$). The time delay between the images can also cause their redshifts to differ because of cosmological expansion. This Differential Expansion Effect is likely to be small. Using a simple model, we derive these effects from first principles.

One application would be to the Bullet Cluster, whose large tangential velocity may be inconsistent with the $\Lambda CDM$ paradigm. This velocity can be estimated with complicated hydrodynamic models. Uncertainties with such models can be avoided using the MCE. We argue that the MCE should be observable with ALMA. 

However, such measurements can be corrupted by the DME if typical spiral galaxies are used as sources. Fortunately, we find that if detailed spectral line profiles were available, then the DME and MCE could be distinguished. It might also be feasible to calculate how much the DME should affect the mean redshift of each image. Resolved observations of the source would be required to do this accurately.

%We argue that the tangential velocity (instead of the hydrodynamical shock speed) of such high-z clusters measure by the MCE can be corrupted by the DME.  

The DME is of order the source angular size divided by the Einstein radius times the redshift variation across the source. Thus, it mostly affects nearly edge-on spiral galaxies in certain orientations. This suggests that observers should reduce the DME by careful choice of target, a possibility we discuss in some detail.

\end{abstract}


%\newpage

\section{Introduction}
 
%The $\Lambda$CDM paradigm \citep{LCDM_Proposal} appears to describe large scale structure and cosmology very well (\citet{Planck_Cosmological_Parameters}; \citet{Matter_Power_Spectrum}; \citet{Supernova_Cosmology_Project}). However, there are some observations which appear difficult to reconcile with it (\citet{Pawlowski_2014}; \citet{VPOS_Proper_Motions}; \citet{Andromeda_Satellite_Plane}; \citet{MOND_Review} and references therein). One such case is the collision speed of the Bullet Cluster, 1E0657-56 \citep{Tucker_1995}.

The standard $\Lambda$CDM paradigm \citep{LCDM_Proposal} still faces many challenges in reproducing galaxy scale observations \citep[for a recent review, see][]{MOND_Review}. Particularly problematic is the anisotropic distribution of satellites around Local Group galaxies, a question recently revisited in detail \citep{Pawlowski_2014}. A different analysis focusing on Andromeda came to similar conclusions \citep{Ibata_2014}. The relevant observations for the Milky Way \citep{VPOS_Proper_Motions} and Andromeda \citep{Andromeda_Satellite_Plane} are difficult to repeat outside the Local Group because of the need to obtain 3D positions and velocities.

On a larger scale,  \citet{Cai_2014} found that the collision speed distribution of interacting galaxy clusters can be quite sensitive to the underlying law of gravitation. Thus, the high collision speed of the components of the Bullet Cluster 1E0657-56 \citep{Tucker_1995} has been argued in favour of modified gravity \citep{Katz_2013}. However, this speed is not directly measured as the collision is mostly in the plane of the sky. Instead, the speed is estimated using simulations of the shock generated in the gas by the collision. The separation of the DM and gas \citep{Dark_Matter_Proof} also plays an important role - there is less gas drag at lower speeds, so the separation is generally reduced. 

A collision speed close to 3000 km/s is thought to be required to explain the observed properties of the Bullet Cluster \citep{Mastropietro_Burkert_2008}. For the inferred masses of the components \citep{Clowe_2004}, this appears difficult to reconcile with $\Lambda$CDM \citep{Thompson_Nagamine_2012}. This work suggested that a cosmological simulation requires a co-moving volume of $(4.48 h^{-1} \text{Gpc})^3$ to see an analogue to the Bullet Cluster. 

The recent work of \citet{Lage_Farrar_2014} finds a similar collision speed but suggests a higher mass for the Bullet Cluster's components. While higher mass objects are likely to collide faster, such heavy clusters are rare in cosmological simulations. For example, their own unpublished work \citep{Lage_Farrar_2014_Arxiv} based on the Horizon Run N-body simulation \citep{Kim_2009} showed how there were only 7 cluster pairs with masses comparable to their higher estimate for the Bullet Cluster mass. Because a larger volume of $(6.59 h^{-1} \text{Gpc})^3$ was used in the simulation and $\left( \frac{6.59}{4.48} \right)^3\approx 3$, this result is not very surprising in light of previous works.

However, some recent unpublished studies have raised the probability estimate of observing a galaxy cluster merger with properties comparable to the Bullet Cluster. \citet{Bouillot_2014} used a larger box size of $(21 h^{-1} \text{Gpc})^3$, using the DEUS-FUR simulation. \citet{Thompson_Nagamine_2014} took issue with the Friends of Friends algorithm \citep{Davis_1985} long used to search outputs of N-body simulations for analogues to the Bullet Cluster. After switching to the recently developed ROCKSTAR algorithm \citep{Behroozi_2013}, the rate of occurrence of analogues to the Bullet Cluster increased by a factor of $\sim$100. Despite this, \citet{Thompson_Nagamine_2014} quoted the probability of a collision as fast as the observed one as only 1 in 2170, which is still fairly small.

Moreover, a few other massive colliding clusters with high infall velocities have been discovered in the last few years \citep{Gomez_2012, Menanteau_2012, Molnar_2013_Abell}. The El Gordo Cluster (ACT-CL J0102-4915) may be particularly problematic due to its combination of high redshift \citep[z = 0.87,][]{Menanteau_2012}, high mass \citep{Jee_2014} and high inferred collision speed \citep{Molnar_2014}.

A detailed analysis of how likely it is that observers would have seen interacting clusters with the observed properties is still lacking. One would need to account for incomplete sky coverage and perhaps faster collisions being easier to discover due to greater shock heating of the gas. A key input to any such analysis must be the collision speeds of the components. This work focuses on measuring cluster motions more accurately.

\citet{Molnar_2013_Abell} argue that inferring collision speeds from observations of the shock can be non-trivial just due to projection effects, let alone other complexities of baryonic physics. To see if there is any tension with the $\Lambda CDM$ model, the collision speeds should be determined in a more direct way. Ultimately, we would like to determine the proper motions of colliding clusters.

%Despite the success of the $\Lambda$CDM paradigm on large scales, there remain observations which are hard to reconcile with it. One such case is the collision velocity of the Bullet Cluster, 1E0657-56. The separation of the dark matter - as inferred from weak gravitational lensing - and the gas, combined with its morphology, appear difficult to reproduce in non-cosmological simulations without collision velocities exceeding 3000 km/s (e.g. \cite{mastropietro_Burkert_2008} and \cite{Lage_Farrar_2014}). The latter (more recent) work appears consistent with $\Lambda$CDM whereas the former does not. This is mainly due to a substantial upwards revision of the mass of the Bullet cluster's components.

%Such simulations are complicated and would likely benefit greatly from independent constraints on the velocities of the components. Because the collision appears to mostly be in the plane of the sky, this is not feasible by measuring redshifts.

%However, this conclusion is based on comparison of observations with simulations of complicated gas hydrodynamical processes. Consequently, it is desirable to find independent methods of determining the collision velocity.

 %It would be even better to determine the proper motion of the dark matter halos rather than the motion of the gas, as the former is unaffected by hard-to-model gas drag and so offers better insight into the gravitational forces acting on the system. 

Although not feasible by traditional methods, such motions may be inferred using the Moving Cluster Effect \citep[MCE,][]{Birkinshaw_Gull_1983}. As derived later, this effect relies on the gravitational potential of an object being time-dependent due to its motion. Consequently, if a source behind the object were multiply imaged, the images would have slightly different redshifts. Moreover, as the dark matter generally outweighs the gas on cluster scales \citep{Blaksley_Bonamente_2009}, the MCE is mostly sensitive to the motion of the dark matter. This is simpler to model than the gas, making the results easier to compare with simulations.

The MCE would likely be around 1 km/s for the Bullet Cluster \citep{Molnar_2013}. The Effect may be searched for using Cosmic Microwave Background (CMB) photons \citep[e.g. ][]{Cai_2010}. However, as noted by those authors, temperature anisotropies in the CMB make it difficult to spot such a signal around an individual object. Thus, we focus instead on using a multiply imaged background galaxy as the source. Spectral features in this galaxy could be used to determine the redshifts of its multiple images. 

We consider the feasibility of obtaining measurements of the required accuracy using ALMA in Section \ref{Observations}. Measuring this Effect seems to be within our reach. One might instead conduct the observations in the visible/near-IR with large spectroscopic instruments such as the TMT and E-ELT. %These facilities should also be able to obtain sufficiently accurate measurements. 

Therefore, it is important to consider other effects that might also cause the redshifts of double images to differ. Perhaps the most important such mechanism is what we term the $\text{Differential}$ $\text{Magnification}$ $\text{Effect}$ $\text{(DME)}$. This depends on details of the source. If this is a rotating disk galaxy not viewed face-on, then different parts of the source have different radial velocities and hence redshifts. 

The lens magnifies the source non-uniformly. The exact way in which this occurs is different for each image. Consequently, the intensity-weighted mean redshift of the images is usually different.

If one could perform Integral Field Spectroscopy of the source galaxy accurate to $\sim$1 km/s, then one would simply need to compare the redshift of the same part of the galaxy between the two images. By focusing on a small part of the galaxy, the DME would be greatly reduced. However, this will be a challenging observational goal. The high spectral accuracy demanded by MCE measurements means the source will likely be spatially unresolved in the near future.

Assuming this to be the case, we determine the order of magnitude of the DME for a typical disk galaxy. We find that it may well be significant. Thus, we explore exactly how it affects the profiles of individual spectral lines. The Effect is quite different to the MCE, which simply shifts each line. This may provide a way to correct for the DME and also to verify that a redshift difference is indeed caused by the MCE. Without spectra detailed enough to see such small differences between line profiles, it might still be possible to calculate the DME, though the determination would be less secure.

The additional complications and uncertainty introduced by trying to correct for the DME necessitate a discussion on how it may be reduced. Aside from the obvious steps of using ellipticals/face-on spirals and smaller $-$ likely slower-rotating $-$ galaxies, an important factor to consider is how much the magnification varies across the source. The larger the variation, the larger the DME. 

For this reason, an edge-on fast-rotating spiral galaxy might still be a good target if it is oriented so the magnification is nearly constant across the image. At the other extreme, the magnification varies rapidly near a caustic. Therefore, caustic images are likely to be strongly affected by the DME \citep{Molnar_2013}.

We emphasise the need to model both the redshift structure of the source and the deflection map of the lens when trying to use precise lensed image redshifts to determine the tangential motion of the lens.

%\newpage

\section{The Moving Cluster Effect in Standard and Non-Standard Gravity}

\subsection{The Lensing Geometry}
\label{Geometry}

\begin{figure}
	\centering
	\includegraphics [width = 8cm] {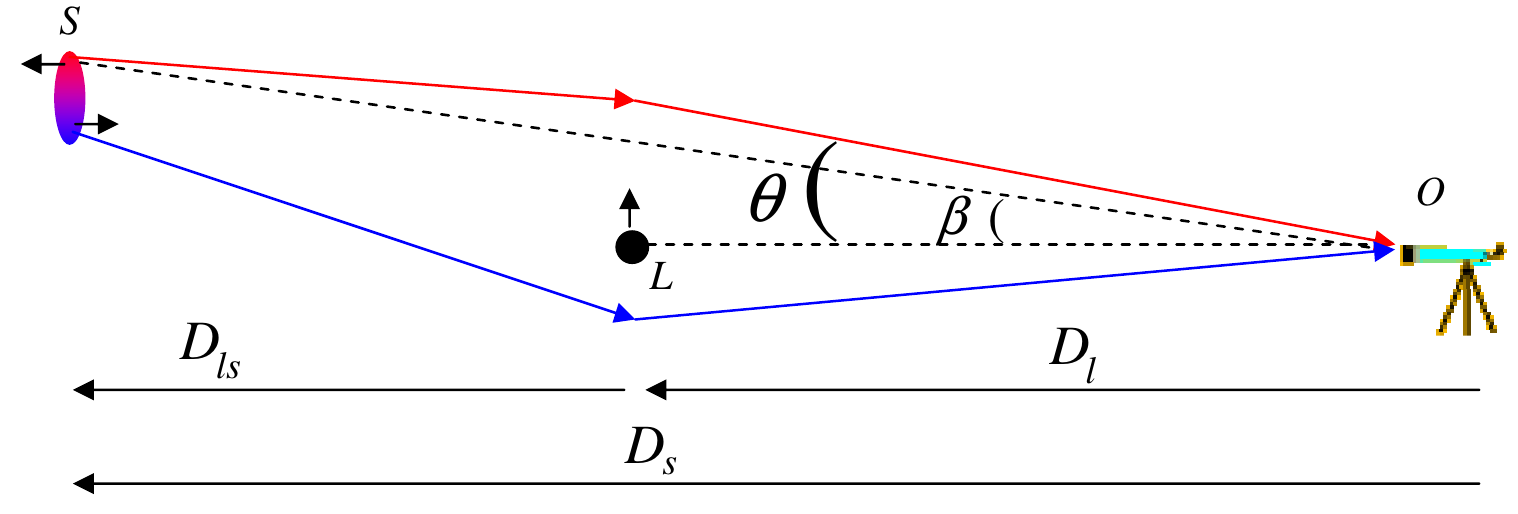}
	\caption{The lensing geometry is depicted here. Upper photon trajectory $=$ primary image (same side as unlensed source), lower trajectory $=$ secondary image. Relevant distances are indicated at bottom. The lens is treated as a point mass moving transversely to the viewing direction. The source is an extended disk galaxy. There is a redshift gradient across it due to rotation.}
	\label{Lensing_Geometry}
\end{figure}

\text{Figure} \ref{Lensing_Geometry} illustrates the basic geometry that will be considered here. Because we are mostly concerned with angles on the sky, the distances relevant to us are the angular diameter distances to the lens and source ($D_l$ and $D_s$). Also important is the angular diameter distance to the source galaxy as perceived by an astronomer at the lens, measured at the epoch that the lens is currently observed at ($z_l$). This last distance we denote ${D}_{ls}$.

%The primary image is on the same side as the unlensed source. The secondary image is inverted

%\footnote{Some workers have an additional $O$ subscript in some of these distances, which we omit for clarity. A distance must be between two points and if only one is specified, then the other is Earth.} 

\subsection{Including Lens Motion}

Suppose that the lens moves transversely to $\bmath{OL}$. Thus, one of the light paths gets `stretched' while the other gets `squeezed', leading to a redshift difference between the images. To calculate this effect, we make the thin-lens or triangle approximation whereby each photon trajectory is treated as two straight lines. In this case, the light arrival time surface is generally given by \citep{Kovner_1990}
\begin{eqnarray}
	cT({\bmath{\theta}})  = constant + \frac{D_l D_s}{2 D_{ls}} ({\bmath{\theta}} - {\bmath{\beta}})^2 - \Psi(D_l {\bmath{\theta}} - \bmath{x_l})
	\label{Time_of_flight}
\end{eqnarray}
 
This consists of a geometric part and a relativistic part due to the lensing potential $\Psi$. We assume that $\Psi$ depends only on position relative to the lens, which is located at $\bmath{x_l}$ relative to some reference point in the lens plane. 

The path length $cT({\bmath{\theta}})$ can be thought of as a function of the lens plane position ${\bmath{\theta}}$ hit by a ray from the source. The actual rays for the images are at the extrema of this function (Fermat's Principle). 

We briefly describe how the geometric part of Equation \ref{Time_of_flight} is derived. Each section of a hypothetical undeviated photon trajectory can be mapped onto a part of the actual trajectory. The latter is slightly longer as there is an extra factor of the secant of the angle between them. This is expanded at second order, as the angle is small. The angle is $\left( \theta - \beta \right)$ for the stretch $LO$ while for $OS$, it is $\left( \theta - \beta \right) \frac{D_l}{D_{ls}} \frac{1 + z_l}{1 + z_s}$. The last factor arises because photons emitted in the same direction gradually get further apart due to cosmic expansion. Thus, photons emitted in different directions end up more widely separated than in a static Universe.

Equation \ref{Time_of_flight} follows most naturally if using co-moving distances, which can be added simply. This fact leads to the very useful relation between angular diameter distances
\begin{eqnarray}
	\left(1 + z_l \right) D_l ~+~ \left( 1 + z_s \right) D_{ls} = \left( 1 + z_s \right) D_{s}
\end{eqnarray}	

If the lens moves in the transverse direction, then $\bmath{x_l}$ changes and so the lensing potential $\Psi$ changes at every point in the lens plane. Thus, transverse motion of the lens would cause the arrival time to change according to 
%For a fixed source position ${\bmath{\theta}}$, as 
\begin{eqnarray}
	c\dot{T}   = \dot{\bmath{x_l}} \cdot \bmath{\nabla} \Psi  \equiv  -\bmath{v_t} \cdot \bmath{\alpha} 
\end{eqnarray}

$\bmath{\alpha} \equiv -\bmath{\nabla} \Psi$ is the unreduced (true) deflection angle and $\bmath{v_t} \equiv \dot{\bmath{x_l}} $ is the transverse velocity of the lens (note time here refers to that measured by a clock at the lens). 

The rate of increase of the path length $c\dot{T}$ is equivalent to a shift of the intrinsic spectrum of the source. While the source's intrinsic spectrum can't be directly measured, the relative spectra of the images in a multiple-image system can. Images 1 and 2 would have a relative redshift velocity 
%$\delta V_r$ with 
%relative proper motion of the lens projected as the transverse velocity in the lens plane
\begin{eqnarray}
	\delta V_r \equiv V_1 - V_2  = -\bmath{v_t} \cdot (\bmath{\alpha_1} - \bmath{\alpha_2})
	\label{MCE_vector_form}
\end{eqnarray}

where $\bmath{\alpha_1} - \bmath{\alpha_2}$ is the relative deflection angle between the images. The observed angular separation between the images is $\frac{D_{ls}}{D_s}$ times as much. This allows the MCE to be calculated without knowing what the deflection angles are, as long as one is sure of the identification of the double image and the distances to the lens and source. % $\alpha$ is the true deflection angle of an image.

So far, we have used time measured by a clock at the lens. Using one on Earth instead would introduce a factor of $(1+z_l)$ to the time delay. Putting it in, we should think of $\bmath{v_t}$ as the transverse peculiar velocity in co-moving coordinates. This is the co-moving lens distance times the relative proper motion of the lens with respect to that of the source.\footnote{Physically, it would be the peculiar velocity of the lens in the direction orthogonal to our line of sight, in the absence of peculiar motions of either observer or source.} This takes account of transverse motions of the observer and the source. In Section \ref{Observer_Source_Motion}, we show that such motions affect image redshifts much less than motion of the lens.

{\it For multiple lens planes, one would simply add the redshift differences due to each plane.}

The above derivation is a property of metric theories of gravity. Hence, it is independent of details of the theory, something we now show. Consider a static Universe with no observer-source relative motion. Use a reference frame moving with the lens, so the source and observer both appear to move at $-\dot{\bmath{x_l}}$. In general, there is a Doppler shift for a photon emitted by the source as perceived at the lens. A similar effect arises between lens and observer. The shifts cancel if the photon is not deflected by the lens (if the source emits a photon `backward', then the observer `ploughs into' the photon). 

However, if the photon is deflected, then there is a net frequency shift between source and observer. This shift is different for another photon which gets deflected by a different amount. Therefore, the \textbf{difference} in deflection angles determines the redshift difference between the photons. If observers could be sure the photons initially had the same energy, then the redshift difference could be directly measured, thus constraining $\dot{\bmath{x_l}}$.

For small deflections by a non-relativistic lens, the result in an expanding Universe is exactly the same as in a static one, once all the angles have been properly accounted for.

%The above derivation is a property of metric theories of gravity. Hence, it is independent of details of the theory. One can understand the above calculation in the following classical dynamics analogy. Considering sending two ping-pong balls colliding with two parallel walls at grazing angles so that they are deflected by the angles  $\bmath{\alpha_1}$ and $\bmath{\alpha_2}$ respectively. Now let both walls move in the same transverse direction with a velocity $\dot{\bmath{x_l}}$, then the two ping-pong balls will be redshifted by an amount $-\dot{\bmath{x_l}} \cdot \bmath{\alpha_1}$ and $-\dot{\bmath{x_l}} \cdot  \bmath{\alpha_2}$ respectively.

\subsection{Point mass lenses}

The unlensed source and images all lie along a line, so we only consider positions along this line. For a point mass lens, one can generically say that a ray of light with impact parameter $D_l \theta$ is deflected by
\begin{eqnarray}
	\alpha = -\frac{4G\widetilde{M}}{c^2 D_l \theta }
	\label{Deflection_angle}
\end{eqnarray}

$\widetilde{M}$ is the `equivalent lensing mass' in alternative gravity. Different gravity theories require different amount of real mass $M$ to produce the observed equivalent mass $\widetilde{M}$, the Einstein radius $\theta_E$ or the deflection angles $\alpha_{1,2}$. In general, $\widetilde{M}$ depends on position in modified gravity theories, even with a point mass. For simplicity, we neglect this. 

Combining Equations \ref{MCE_vector_form} and \ref{Deflection_angle}, we get that
\begin{eqnarray}
	%V_1 - V_2 =   \left( \frac{\theta_E^2}{\theta_1} - \frac{\theta_E^2}{\theta_2} \right) v_t ~~~~~ \text{where     } \theta_E \equiv \frac{4G \widetilde{M} D_{ls}}{ c^2 D_s D_l}
	V_1 - V_2 = \frac{4G\widetilde{M}v_t}{c^2 D_l} \left( \frac{1}{\theta_1} - \frac{1}{\theta_2} \right)
	\label{Delta_v_r}
\end{eqnarray}

Note that the signs of $\theta_1$ and $\theta_2$ are opposite because the images are on opposite sides of the lens (Equation \ref{Image_positions}).

$v_t$ is the component of the transverse velocity $\bmath{v_t}$  projected along the line connecting images 1 and 2. If the lens proper motion is orthogonal to the image separation, then $v_t$ would be zero. 

Noting that deflecting a photon at the lens only affects part of its trajectory, we get the classical lens equation
\begin{eqnarray}
	\beta &=& \theta -\frac{{{D}_{ls}}}{{{D}_{s}}}\frac{4G\widetilde{M}}{{{c}^{2}}{{D}_{l}}\theta }\\
	&\equiv& \theta - \frac{{\theta_E}^{2}}{\theta} ~~~\text{where}~	{{\theta }_{E}} \equiv \sqrt{\frac{4G\widetilde{M}}{{{c}^{2}}}\frac{{{D}_{ls}}}{{{D}_{l}}{{D}_{s}}}}
	\label{Einstein_Radius}	
\end{eqnarray}

The Einstein radius $\theta_E$ defines a typical angular scale for the problem. It will be convenient to use this to normalise all relevant angles. Thus, we let
\begin{eqnarray}
	u \equiv \frac{\beta}{\theta_E}   ~~~~~   \text{and}   ~~~~~  y \equiv \frac{\theta}{\theta_E}
\end{eqnarray}

Images are formed where 
\begin{eqnarray}
	y = \frac{1}{2}\left( u\pm \sqrt{u^2 + 4} \right)
	\label{Image_positions}
\end{eqnarray}

For later use, we note that the magnification of a small part of the source located at an unlensed angular position of $\beta \equiv  u \theta_E$ relative to the lens is given by
\begin{eqnarray}
	A &=& \left| \frac{\theta }{\beta }\frac{\partial \theta }{\partial \beta } \right| \\
	&=& \frac{1}{2}\left( \frac{{{u}^{2}}+2}{u\sqrt{{{u}^{2}}+4}}\pm 1 \right)
	\label{A}
\end{eqnarray}

The result follows from the surface brightness of an unlensed source and a lensed one being equal (Liouville's Theorem). Thus, $A$ is the Jacobian of the mapping between where objects appear on the sky and where they would without a lens. The modulus signs are needed because otherwise $A < 0$ for the secondary image (indicating that it is inverted). 

The secondary image becomes very faint if $u \gg 1$ (in this case, $A \sim \frac{2}{u^4}$). It is difficult to find a source with $u \ll 1$ as this corresponds to a very small part of the sky. Thus, any source used for MCE measurements will very likely have $u \sim$1. We assume this is the case.

\section{Causes Of Differences Between Multiple Image Redshifts}
\label{Redshift_splitting}

Before deriving the DME, we briefly consider a few factors that can affect redshift differences between double images of a source strongly lensed by the Bullet Cluster.

\subsection{Lens Motion}
\label{Lens_Motion}

The MCE is maximal for a source displaced from the lens in the direction of its proper motion. This direction can be determined from images of the shock fairly easily. Moreover, observers should select targets to maximise the MCE. Thus, we assume the double images are indeed separated along the direction of motion of the lens. Otherwise, the MCE is reduced by the cosine of the angle between them (Equation \ref{MCE_vector_form}). 

With these assumptions, we combine Equations \ref{Delta_v_r} and \ref{Image_positions} to get that the difference in redshift velocity between the two images is 
\begin{eqnarray}
	{{\left. \Delta \overline{v_r} \right|}_{MCE}} = \frac{2{v_t}\sqrt{GM\left( {{u}^{2}}+4 \right)}}{c}\sqrt{\frac{{{D}_{s}}}{{{D}_{ls}}{{D}_{l}}}}
	%& {v_t}\left( \frac{4GM}{{{D}_{l}}{{\theta }_{1}}{{c}^{2}}}-\frac{4GM}{{{D}_{l}}{{\theta }_{2}}{{c}^{2}}} \right)\\
	\label{MCE}
\end{eqnarray}

Using parameters appropriate to the Bullet Cluster (Table \ref{Inputs}), the effect is around 1 km/s.

\subsection{Source \& Observer Motion}
\label{Observer_Source_Motion}

The peculiar velocity (w.r.t. the CMB) of the Sun is well-known \citep[369 km/s,][]{Planck_CMB_Dipole} and could be included in a more careful analysis. We do not include it as we only seek a rough idea of the magnitude of the MCE. This won't be substantially affected by observer motion as this is much slower than that of the lens ($\sim$3000km/s).

More problematic may be the unknown peculiar velocity of the source. Treating the Local Group peculiar velocity ($\sim$630 km/s) as typical for galaxies, the Bullet Cluster transverse motion likely greatly exceeds the source's peculiar velocity. In this case, only the component of this velocity parallel to the lens transverse motion has much effect on image redshifts, leading to a factor of $\frac{1}{2}$ on average.\footnote{For an angle between a fixed vector and another statistically isotropic one, $\langle \left| \cos \theta \right| \rangle = \frac{1}{2}$.} 

Another factor of $\frac{1 + z_l}{1 + z_s} \frac{D_l}{D_{s}} \approx \frac{1}{4}$ arises due to the geometry of the situation and cosmic expansion. Moreover, typical peculiar velocities were smaller long ago. Supposing they were $\frac{1}{2}$ as much as today at $z_s = 1.5$, we see that source motion can't affect the inferred lens velocity much more than $\sim$50 km/s. This effect can be reduced by observing more than one double image pair. However, we consider the accuracy with even just one well-observed pair sufficient.

Thus, we ignore any motion other than that of the lens. We note that it might be good to avoid source galaxies which are interacting, as their peculiar velocities might be higher.

\subsection{Cosmological Expansion}

A redshift difference between the images can also arise because the time of flight of photons emitted by the source is different depending on which path they took to get to Earth. As the photons for both images arrive simultaneously, the photons for one image must have been emitted earlier than for the other. Thus, in an expanding Universe, one of the images will have a higher redshift. Due to both a longer path length and a stronger gravitational field along the path, this image is the secondary (on the opposite side of the lens as the unlensed source would appear $-$ see Figure \ref{Lensing_Geometry}). We term this the Differential Expansion Effect (DEE).

A quick way to estimate the DEE is to assume that cosmological distances like $D_L$ are usually $\sim$$\frac{c}{H}$. The extra path length $\sim$$D_L \theta^2$. The effect of the difference in gravitational time delays can crudely be approximated as equal to that due to different geometric path lengths. 

The DEE expressed as a redshift $ = H \Delta t \approx \theta^2$. Meanwhile, the MCE $\sim$$\frac{v}{c} \theta$. Assuming a velocity of 3000 km/s and an image separation of 20'', we see that the MCE is $\sim$50 times larger than the DEE. Thus, we calculate the DEE more precisely. 

We first consider just the difference in time of flight (`delay') due to different strengths of gravity along the two possible photon paths \citep{Shapiro_1964}. The relative Shapiro delay between the images is 
\begin{eqnarray}
	\Delta t_l = \frac{4GM}{c^3} Ln\left( \frac{b_2}{b_1} \right)~\text{,     } D_s, D_{ls} \gg b_{1, 2}
	\label{Shapiro_delay}
\end{eqnarray}

The impact parameters of the rays are $b_{1,2}$. This result is valid if $b$ is much larger than the Schwarzschild radius of the lens, so the rays are only weakly deflected. The ray with smaller $b$ is delayed more as it goes deeper into the lens' gravitational potential well. It also has a longer geometric path length (it forms the secondary image in Figure \ref{Lensing_Geometry}).

Note this is the time delay at the lens. In reality, both photons must reach Earth now, so the time of emission at the source must have been different. This requires an extra factor of the relative rates of a clock at the lens and at the source, $\frac{1 + z_l}{1 + z_s}$. 

We then combined this Shapiro delay with the geometric path difference between the trajectories. Thus, the difference in time of emission required for photons traversing the two trajectories to reach Earth simultaneously is
\begin{samepage}
\begin{eqnarray*}
	\Delta t_s =
\end{eqnarray*}
\begin{eqnarray}
	\frac{1 + z_l}{1 + z_s} \frac{2GM}{c^3} \left[ u\sqrt{u^2 + 4} ~+~ 2Ln\left( \frac{u^2 + u\sqrt{u^2 + 4} + 2}{2} \right) \right]
	\label{Time_delay}
	%&u \equiv \frac{\beta}{\theta_E}
\end{eqnarray}
\end{samepage}

We used Equation \ref{Image_positions} to relate the source position $u$ to the positions of its images. A correction for cosmological time dilation was also applied. 

The change in redshift is given by the fractional difference in the scale factor of the Universe at the time of emission of the photons.
\begin{eqnarray}
	\Delta z = H( z_s) \Delta t_s
%H\left( z_s \right) = H_0 \sqrt(Omega_{M,0} (1 + z_s)^3 + Omega_{\Lambda,0}) 	
\end{eqnarray}

Using realistic parameters (Table \ref{Inputs} and $u \approx 1$), the DEE $\sim$1 m/s. In Section \ref{Lens_Motion}, we showed that the MCE is $\sim$1000 times larger, allowing us to neglect the DEE. 

Even with more accurate instruments, a very large number of double image pairs would need to be observed before the random noise from source peculiar motions was reduced below such a small level. Thus, the DEE won't be important in the foreseeable future. An exception might possibly arise if the source galaxy peculiar motion could be estimated based on properties of galaxies near it.

%We mention briefly that if observers did try to detect the DEE, then the necessary measurements would likely need to be done over quite a long period. In this case, the `redshift drift' of distant objects due to cosmological expansion over the observing period might need to be accounted for \citep{McVittie_1962}. This is especially true if the observations for each image pa

\section{Derivation of the Differential Magnification Effect for Unresolved Images}

\begin{figure}
	\centering
		\includegraphics [width = 8cm] {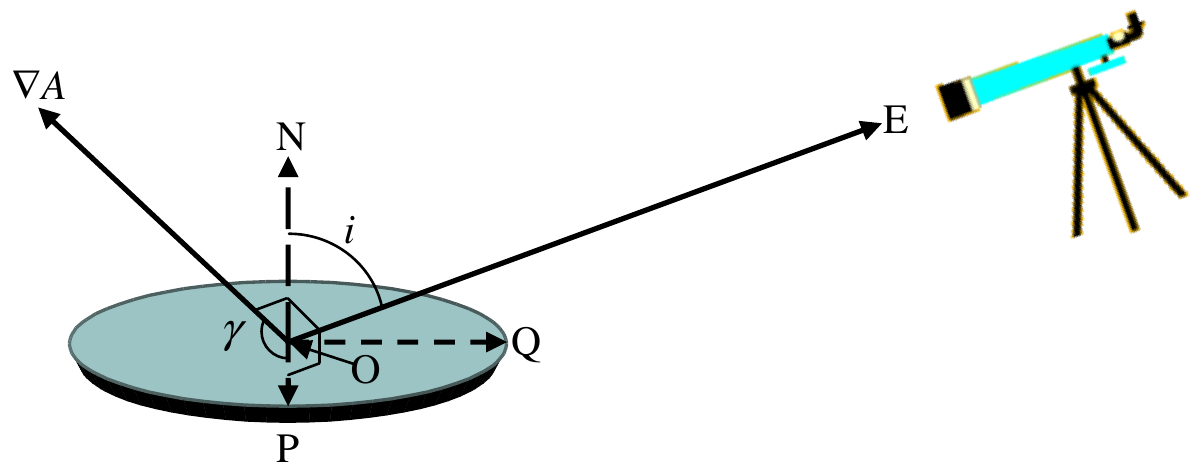}
	\caption{The observing geometry is shown here. The source galaxy has centre O and normal to its plane $\bmath{ON}$. Earth is towards $\bmath{OE}$, so the galaxy's inclination to the sky plane is $i$. $\bmath{OQ}$ and $\bmath{OP}$ are in the galaxy's plane and orthogonal to each other, with $\bmath{OQ}$ as closely aligned with $\bmath{OE}$ as possible. Thus, $\bmath{OP}$ and $\bmath{OE}$ are orthogonal. $\bmath{\nabla} A$ is directed within the source plane, so must also be orthogonal to $\bmath{OE}$. $\bmath{\nabla} A$ is at an angle $\gamma$ to $\bmath{OP}$. The source is parameterised using cylindrical polar co-ordinates ($r$, $\phi$), with centre $O$ and initial direction ($\phi = 0$) along $\bmath{OQ}$.}
	\label{Source_geometry}
\end{figure}

%The angle $\phi$ as measured from $\bmath{OQ}$ and the distance $r$ is used to parameterise the galaxy, as is the quantity $r$ which measures distance from $O$ (neither shown).

%We now decompose the source galaxy into small portions over which the magnification $A$ is constant. This is done in Figure \ref{Source_geometry}. 

The effects mentioned in Section \ref{Redshift_splitting} cause the frequencies of identical photons emitted in different directions to end up different when measured at the Earth. The DME does not do this. It is merely an observational artefact due to inability to simultaneously resolve the images and take highly accurate spectra of them. This causes parts of the source with different redshifts to get blended together in the spectra. The precise way in which this blending occurs is different between the images.

We assume the spectra are integrated over the entirety of each image. The source is modelled as a typical spiral galaxy with exponential surface density profile and a realistic rotation curve. The lens is treated as a point mass. The parameters considered (Table \ref{Inputs}) are designed with the Bullet Cluster \citep{Tucker_1995} in mind. 

The basic idea is that spatially unresolved spectra can determine the intensity-weighted mean redshift of each image. This may be affected by rotation of the source galaxy. The effect isn't reliant on an expanding Universe, so it will be simplest to think of the Universe as static for the remainder of this section.

The mean redshift velocity of each image is given by
\begin{eqnarray}
	\overline{v_r} \equiv \frac{\int_{\text{Image}}{A\Sigma {{v}_{r}}}~dS}{\int_{\text{Image}}{A\Sigma }~dS}
	\label{Governing_Equation}
\end{eqnarray}

The integrals are over area elements of the source $S$. This is treated as a disk with surface density
%$S$ is assumed inclined by only $30^\circ$ to the plane of the sky, but with the major axis of the image pointing directly away from the mass. Relaxing the last assumption would reduce the magnitude of the effect, something we compensated for by lowering its inclination.
\begin{eqnarray}
	\Sigma = \Sigma_0 ~ {e}^{-\frac{r}{r_d}}
	\label{Surface_density}
\end{eqnarray}

The magnification $A$ varies little over the source galaxy. This is because $\frac{r_d}{D_s} \ll \theta_E$ (see Table \ref{Inputs}). Thus, a linear approximation to $A$ is sufficient.
\begin{eqnarray}
	A \approx A_0 + \frac{\partial A}{\partial u} du ~~~(A_0 \equiv A ~\text{at centre of source})
\end{eqnarray}

In our model, $A$ varies linearly with position in the source plane, but only in the direction directly away from the (projection of the) lens. In the orthogonal direction, $A$ is independent of position at first order (because $u$ is, and $A$ depends only on $u$).

The geometry of the source is shown in Figure \ref{Source_geometry}. The radial velocity of any part of it is 
\begin{eqnarray}
	v_r = v_c(r) ~ \sin \phi ~ \sin i
\end{eqnarray}

Only the component of $\bmath{\nabla} A$ along $\bmath{OP}$ is important. To see why, suppose that $\bmath{\nabla} A$ was entirely along $\bmath{OQ}$. Reflecting the galaxy about the line $OP$ without altering $\bmath{\nabla} A$ should reverse the DME as this is equivalent to reversing $\bmath{\nabla} A$. However, the radial velocity of every part of the galaxy remains unaltered after the reflection (as $\phi \to \pi - \phi$). Thus, the DME must also remain unaltered. 

Noting that the component of $\bmath{\nabla} A$ along $\bmath{ON}$ is irrelevant for the DME, we see that only the component along $\bmath{OP}$ might be relevant. This component causes the approaching and receding halves of the galaxy to be magnified differently. It will be responsible for the DME. Thus, we assume the lens is located somewhere along the line $\bmath{OP}$, making $\bmath{\nabla} A$ entirely along this direction. The result is then multiplied by $\cos \gamma$.

%BRACKETS AROUND COS(ALPHA) ARE ACTUALLY REQUIRED!!!

The magnitude of the DME is therefore $\propto \cos \gamma ~ \sin i$. Assuming isotropy, all values of $\gamma$ are equally likely. But values of $i$ close to $\frac{\pi}{2}$ are more likely because there are more ways for two vectors to be orthogonal than to be aligned. This means the ratio between the average magnitude of the DME and the maximum it could be is given by the mean of $|\cos \gamma ~ \sin i|$, with $\gamma$ unweighted but a further $\sin{i}$ weighting over $i$.\footnote{Edge-on galaxies are less likely to be detected due to dust obscuration. This makes low values of $i$ - and thus a smaller DME - more likely, for a randomly selected multiple image.} Thus,
\begin{eqnarray}
	\langle|\sin i ~ \cos \gamma|\rangle &=&	\frac{\int\limits_{0}^{\frac{\pi }{2}}{\cos \gamma ~d \gamma}}{\int\limits_{0}^{\frac{\pi }{2}}{d\gamma}}\times \frac{\int\limits_{0}^{\pi }{{{\sin }^{2}}i}~d i}{\int\limits_{0}^{\pi }{\sin i ~d i}} \nonumber \\ 
	&=& \frac{1}{2}
	\label{Angle_averaging}
\end{eqnarray}

The angular separation between the lens and the unlensed source is given by
\begin{eqnarray}
	u = u_0 + \frac{\bmath{r}.\bmath{\widehat{OP}}}{D_s \theta_E} ~~~~~(u_0 \equiv u ~ \text{at centre of source})
\end{eqnarray}

The component of $\bmath{r}$ (measured from the galaxy's centre) along $\bmath{OP}$ is $r \sin \phi$. We have not kept careful track of signs because, for any conceivable orientation, the source galaxy could be rotating in the opposite sense, thereby reversing the DME. We explain which image has a lower redshift due to the DME later. 

%Thus, determining whether the DME adds to the MCE or opposes it for a particular pair of images should be straightforward (given that the motion of the lens can be determined easily as the collision in the Bullet Cluster is viewed nearly side-on).

The difference in $u$ between the centre of the source galaxy and any other point in it is given by 
\begin{eqnarray}
	du = \frac{r ~ \sin \phi}{D_s \theta_E}
\end{eqnarray}

%Thus, increasing the accuracy of the calculation by including the second-order correction to $A$ will introduce a $\sin^2(\phi)$ dependence to it. 
$A_0$ represents a constant magnification across the source. This does not contribute to the numerator in Equation \ref{Governing_Equation} because the radial velocity $v_r \propto \sin \phi$, while $\Sigma$ is independent of $\phi$ due to axisymmetry. Thus, integrating over $\phi$ gives 0.\footnote{This is expected, as the DME doesn't arise if the image is uniformly magnified.} The DME arises when including the first order correction to $A$.

The denominator in Equation \ref{Governing_Equation} is a normalisation for each image (its total intensity).\footnote{What we perceive as the total intensity given $D_s$ and $z_s$, but without correcting for magnification by the lens.} Because the magnification is nearly constant across the source galaxy, we can approximate that $A = A_0$. The first order correction to $A$ would have a $\sin \phi$ dependence, which is irrelevant when integrated over all $\phi$. This further justifies our approximation. Therefore, the denominator in Equation \ref{Governing_Equation} becomes
\begin{eqnarray}
	\int_{\text{Image}}{A \Sigma } ~ dS = \Sigma_0 \pi {r_d}^{2} \left( \frac{{{u}^{2}}+2}{u\sqrt{{{u}^{2}}+4}} \pm 1 \right)
\end{eqnarray}

To understand the sign of the DME, firstly note that regions closer to the lens are magnified more. In our approximation, the numerator in Equation \ref{Governing_Equation} is determined by $\frac{\partial A}{\partial u}$, which is the same for both images (Equation \ref{A}). Therefore, the image with the lower magnification (the secondary image) has a larger $\left| \overline{v_r} \right|$. Thus, if it was known which side of the rotating source was the approaching side, one could determine which image should have a higher mean redshift due solely to the DME.

Including the second order dependence of $A$ on sky position slightly alters the calculations done so far. Because a second order term does not affect the approaching and receding halves of the source galaxy differently, the numerator in Equation \ref{Governing_Equation} is unaltered. But the denominator is affected, because the total intensity of each image may be altered by a second order term. This means that our derivation of the DME has a fractional error which is second order in $\frac{r_d}{D_s \theta_E}$. We consider this acceptable \& proceed to develop a model for the redshift structure of the source. This requires a rotation curve.

\subsection{Model Rotation Curves}

%As we are considering spatially unresolved spectra, the results are most sensitive to the bright inner regions of the source galaxy

It will likely be difficult to directly observe the source galaxy rotation curve $v_c(r)$ as it is very far away. It is also difficult to precisely determine its surface density and thus predict the form of $v_c(r)$. Fortunately, we are considering a disk-integrated spectrum and so the exact shape of $v_c(r)$ will turn out not to be very important once the maximum level $v_{max}$ is fixed. 

To get a rough idea of $v_c(r)$, we take advantage of the tight empirical relation between the forces in rotating disk galaxies as required to sustain their rotation curves and those predicted by Newtonian gravity based on the visible (baryonic) mass \citep[][and references therein]{MOND_Review}. This empirical formalism goes by the name of Modified Newtonian Dynamics \citep{Milgrom_1983}. Regardless of whether it is correct at a fundamental level, it does seem to provide a good empirical way of predicting rotation curves. Here, this is important because measuring the actual rotation curve of the source galaxy would be very challenging. 

The particular empirical relation we adopt follows the work of  \citet{Famaey_Binney_2005}.\footnote{This is the so-called `simple $\mu$-function' in MOND.}
\begin{eqnarray}
	\left( \frac{\left| \bmath{g} \right|}{\left |\bmath{g} \right| + a_0} \right) \bmath{g} = \bmath{g_N}
\end{eqnarray}

$\bmath{g}$ is the true gravitational field strength while $\bmath{g_N}$ is the prediction of Newtonian gravity based on the visible mass. $a_0$ is an acceleration scale ($\approx 1.2 \times {10}^{-10} m$/$s^2$) below which either gravity becomes non$-$Newtonian or dark matter must be considered in addition to the baryons. Thus, the magnitude of the gravitational field is given by
\begin{eqnarray}		
	g = \frac{g_N}{2} + \sqrt{{\left( {\frac{g_N}{2}} \right)}^{2} + g_N a_0}
	\label{True_g}
\end{eqnarray}

%The specific interpolation function that seems to work best is

%\begin{eqnarray}
%		\mu(x) = \frac{x}{1 + x}
%\end{eqnarray}

It is not worthwhile to accurately determine $g_N$ for any particular mass distribution because the actual mass distribution in the source is uncertain. Thus, we approximated $g_N$ using an analytic method. We assumed that, to determine $g_N$ at a particular in-plane location, only material at smaller radii need be considered (we verified that the force from material at larger radii was very small). The Newtonian force at a distance $r$ from the centre of a narrow ring of mass $dM$ and radius $x$ is
\begin{eqnarray}
	g_N \approx \frac{G~dM}{r^2} + \frac{3 G~dM~x^2}{4r^4} ~~~~~(x<r\text{, interior ring})
	\label{Equation_26}
\end{eqnarray}

\begin{figure}
		\includegraphics [width = 8cm] {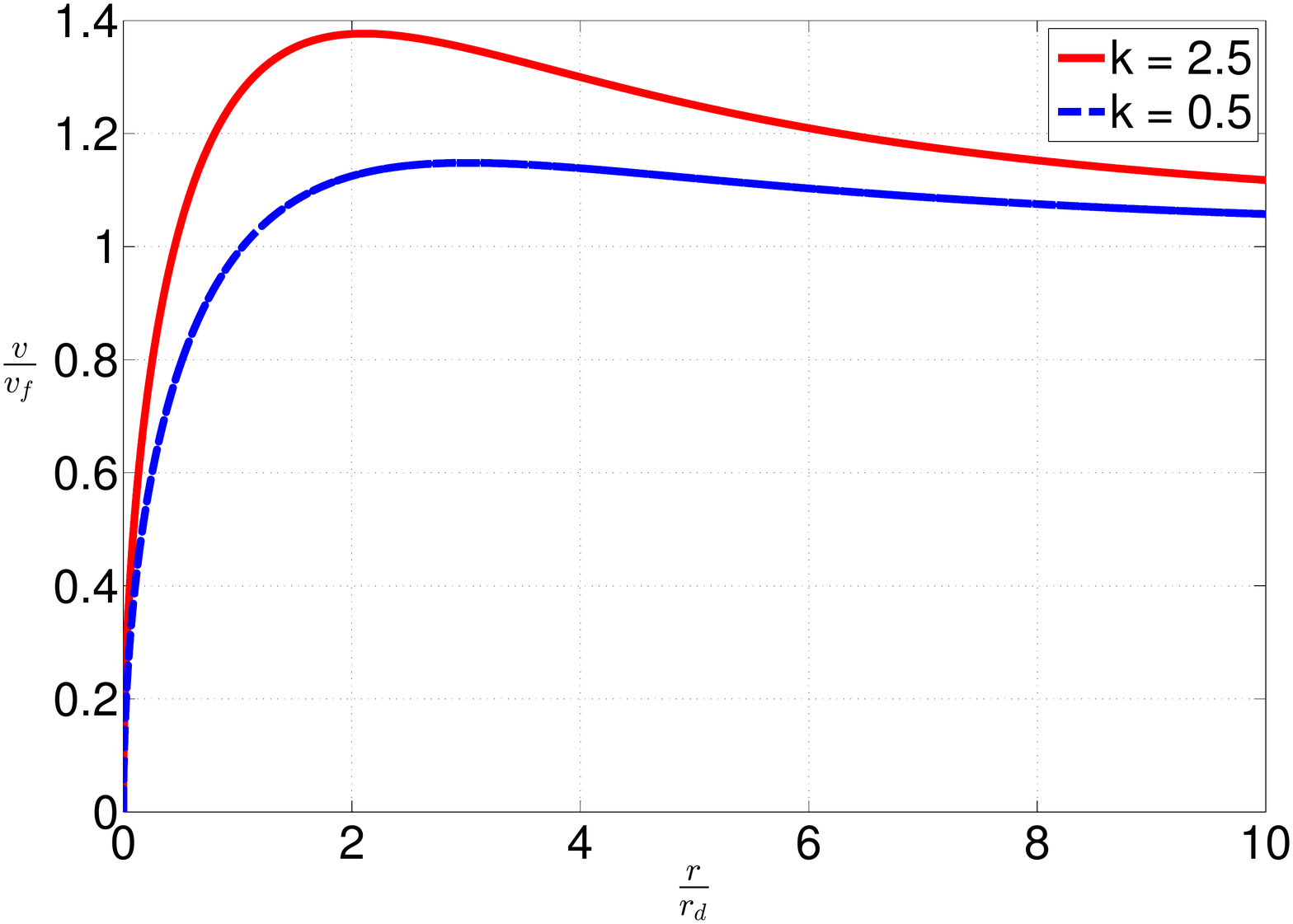}
		\includegraphics [width = 8.5cm] {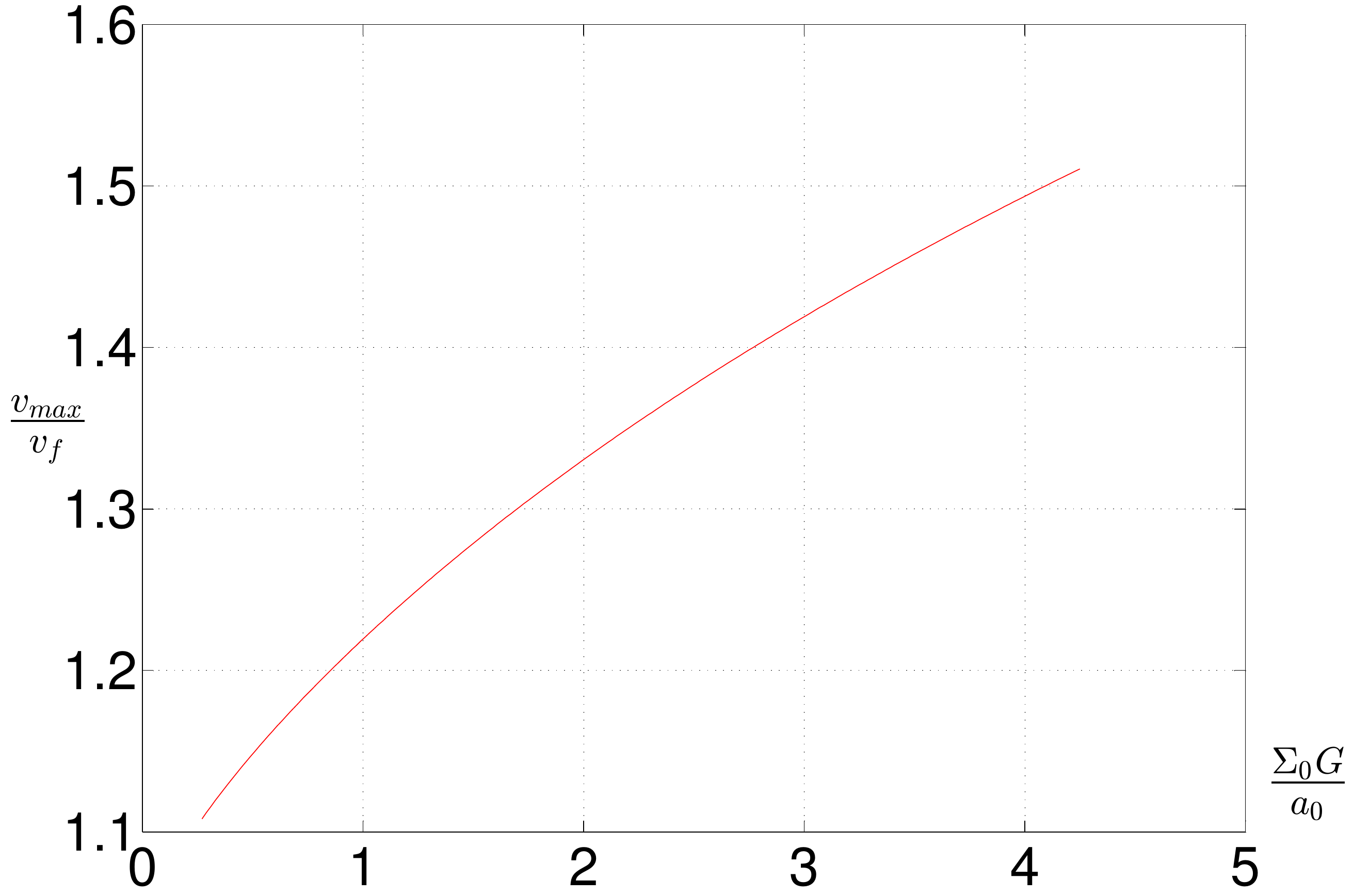}
	\caption{\emph{Top}: Rotation curves resulting from Equations \ref{f} \& \ref{Rotation_Curve}, used in this work. $v_c(r)$ flatlines at $v_f$. The surface density $\Sigma = \Sigma_0 e^{-\frac{r}{r_d}}$. The parameter $k$ controls the shape of the rotation curve ($k \equiv \frac{\Sigma_0 G}{a_0}$). \emph{Bottom}: The ratio of maximum to flatline rotation speed as a function of central surface density.}
	\label{Model_Rotation_Curve}
\end{figure}

%is given by Equation \ref{Surface_density}
%and is proportional to the central surface density (Equation \ref{k})
This is correct at second order in $\frac{x}{r}$. The total force at any point within the disk was found by decomposing the galaxy into a large number of rings with $dM = 2 \pi x ~ dx ~ \Sigma(x)$. We then summed only the forces resulting from interior rings. Therefore, 
\begin{eqnarray}
	g_N &=& \int_{0}^{r}\left( \frac{G~dM}{r^2} + \frac{3 G~dM~x^2}{4r^4} \right)\\
	&=& 2\pi G \Sigma_0 f(\widetilde{r})~~~\text{    where    }\\
	\widetilde{r} &\equiv& \frac{r}{r_d}~~~\text{    and    }\\
	f(\widetilde{r})  &=& \frac{1 - \frac{13}{4} {e}^{-\widetilde{r}}}{\widetilde{r}^2} - \frac{7 {e}^{-\widetilde{r}}}{4\widetilde{r}} + \frac{9 \left( 1 - {e}^{-\widetilde{r}} - \widetilde{r}{e}^{-\widetilde{r}}\right)}{2 \widetilde{r}^4}
	\label{f}
\end{eqnarray}

When obtaining the true value of $g$ from $g_N$, the ratio $\frac{g_N}{a_0}$ is important. Therefore, we introduce a new variable, the dimensionless density $k$. 
\begin{eqnarray}
	k \equiv \frac{G \Sigma_0}{a_0}
	\label{k}
\end{eqnarray}

Typical values for $k$ are order 1. Using the empirical \text{Equation} \ref{True_g} to get $g$ from $g_N$,
\begin{eqnarray}
	\frac{g}{a_0} = \pi k f(\widetilde{r}) + \sqrt{{\left( \pi k f(\widetilde{r}) + 1 \right)}^{2} - 1}
\end{eqnarray}

To get the rotation curve, we equate $g$ with the centripetal acceleration. Thus,
\begin{eqnarray}
		\frac{v_c^2}{\widetilde{r} ~ r_d} &=& \left( \frac{g}{a_0} \right) a_0 \\
		v_f &=& \sqrt[4]{2 \pi k} \sqrt{r_d a_0}
\end{eqnarray}

The rotation curve flatlines at the level $v_f = \sqrt[4]{GMa_0}$, where the total disk mass $M = 2 \pi \Sigma_0 r_d^2$. The shape of the rotation curve is given by
\begin{eqnarray}
		\widetilde{v_c}(\widetilde{r}) \equiv \frac{v_c(\widetilde{r})}{v_f} = \frac{\sqrt{\pi k \widetilde{r} f(\widetilde{r}) + \widetilde{r} \sqrt{\left( {\pi k f(\widetilde{r} ) + 1} \right)^{2} - 1}}}{\sqrt[4]{2 \pi k}}
		\label{Rotation_Curve}
\end{eqnarray}

%We have omitted the normalisation $\Sigma_0$ because this would cancel in Equation \ref{Governing_Equation}. The rotation curve of the galaxy is chosen to be both realistic and result in analytic integrals. We choose a shape appropriate for a HSB galaxy, with a peak at $r \approx r_d$. The rotation curve eventually tends towards a constant value, $v_f$. The parameter $a$ controls the shape of the rotation curve, with higher values giving a larger peak rotation velocity for the same flatline level. We suggest $a$ lies between 1 and 2, values which yield the rotation curves shown in Figure \ref{Rotation_Curve}.

%\begin{eqnarray}
%	v(r)={{v}_{f}}\left( 1-a \left( \frac{1}{a} -\frac{r}{{{r}_{d}}} \right){{e}^{-\frac{r}{{{r}_{d}}}}} \right)
%\end{eqnarray}

%For very low $k$, $v_c(r)$ doesn't peak but slowly rises towards $v_f$.

\subsection{The Final Result}

\begin{samepage}
Combining our results, we get that 
\begin{eqnarray*}
	\left| \overline{v_r} \right| =
\end{eqnarray*}
\begin{eqnarray}		
	\frac{v_f r_d \sin i\cos \gamma }{D_s \theta _E} \frac{\int\limits_{0}^{\infty }{\int\limits_{0}^{2\pi }{\overbrace{{{e}^{-\widetilde{r}}}}^{\propto \Sigma }\widetilde{v_c}(\widetilde{r}){{\widetilde{r}}^{2}}}\overbrace{\frac{4}{{{u}^{2}}{{({{u}^{2}}+4)}^{\frac{3}{2}}}}}^{-\frac{\partial A}{\partial u}}{{\sin }^{2}}\phi ~ d\phi ~ d\widetilde{r}}}{\pi \underbrace{\left( \frac{{{u}^{2}}+2}{u\sqrt{{{u}^{2}}+4}}\pm 1 \right)}_{\propto A}}
	\label{Intermediate_Equation}
\end{eqnarray}
\end{samepage}

%Written in this form, it is easy to see that the DME is of order $\frac{{{v}_{f}} r_d \sin i\cos \alpha }{{{D}_{S}}{{\theta }_{E}}}$. Roughly speaking, this is because t
The magnification $A$ changes by order 1 over an angular distance of $\theta_E$, while the angular radius of the source galaxy $\sim$$\frac{r_d}{D_s}$. Thus, the DME as a fraction of the typical radial velocity of the source is $\sim$$\frac{r_d}{D_s \theta_E}$. The galaxy's typical radial velocity is $v_f \sin i$. Another factor of $\cos (\gamma)$ is needed to account for the lensing geometry. As can be seen from Equation \ref{Intermediate_Equation}, this provides a rough guide to the DME (as $u \sim$1 for a realistic target).

%Thus, the DME is of order $\frac{{{v}_{f}} r_d \sin i\cos \alpha }{{{D}_{S}}{{\theta }_{E}}}$.

An important quantity for the DME is the difference in $\frac{1}{A} \frac{\partial A}{\partial u}$ between the images.
\begin{eqnarray}
	\Delta \left( \frac{1}{A} \frac{\partial A}{\partial u} \right) = \frac{\partial A}{\partial u} \Delta \left( \frac{1}{A} \right) = \frac{4}{\sqrt{u^2 + 4}}
\end{eqnarray}

In Equation \ref{Intermediate_Equation}, the integration over $\phi$ yields $\pi$. The integral over $r$ is not analytic. Thus, we define
\begin{eqnarray}
	I \equiv \int_{0}^{\infty}{{e}^{-\widetilde{r}} \widetilde{v_c}(\widetilde{r})~\widetilde{r}^2} d\widetilde{r}
\end{eqnarray}

Substituting for $\theta_E$ using Equation \ref{Einstein_Radius}, we get that
\begin{eqnarray}
	{{\left. \Delta \overline{v_r} \right|}_{DME}} = \frac{v_f ~ r_d ~ \sin i ~ \cos \gamma ~ I ~ c ~ \sqrt{D_l}}{\sqrt{u^2 + 4} ~ \sqrt{GM{D}_{ls}D_s}}
	\label{DME}
\end{eqnarray}

The integral $I$ depends somewhat on the central surface density in the sense that, for the same $v_f$, the DME is greater at higher $k$. However, the \emph{maximum} rotation speed is very well correlated with the DME. In fact, the ratio $\frac{I}{\widetilde{v}_{max}} = 1.89 \pm 0.02$ for $k = 0.1 \to 5$. As maximum rotation speeds are generally easier to determine than the flatline level, this makes correcting for the DME easier.

If the surface density declines sufficiently slowly with $r$, then the integral $I$ might diverge. This is due to limited validity of a linear approximation to $A$ $-$ a more careful treatment would be required. This might apply to rotating elliptical galaxies with $\rho \propto r^{-4}$. But even if $\Sigma \propto r^{-3}$, the divergence of $I$ is fairly slow. Thus, although the integral would need a cut-off radius, its precise value would not affect the result much. 

A linear approximation to $A$ must break down if $u$ changes by order 1. Thus, a logical cut-off might be the Einstein radius (at the source plane) or the distance between the source and the projected lens. 

If a fibre-fed spectrograph was used or the field of view was otherwise restricted, then this may impose an obvious cut-off. For an inclined disk galaxy, a circular aperture would cover a non-circular region in the disk plane. One might need to take this into account.

%\textbf{If MCE already explained early on, don't repeat.}

%It will be important to compare the difference in $\overline{v_r}$ between the images due to differential magnification with that caused by the Moving Cluster Effect. 

We now decide on realistic parameters to gain a feel for the scale of the DME \& MCE. Ideally, one would like to measure the motion of both components of the Bullet Cluster. However, we choose a mass corresponding roughly with the sub-cluster in the Bullet \citep{Mastropietro_Burkert_2008}. This is because the centre of mass likely has little peculiar velocity as there is little structure on such large scales. Thus, the lower mass component probably moves faster (w.r.t the Hubble flow). 

\begin{figure}
	\centering
		\includegraphics [width = 8.5cm] {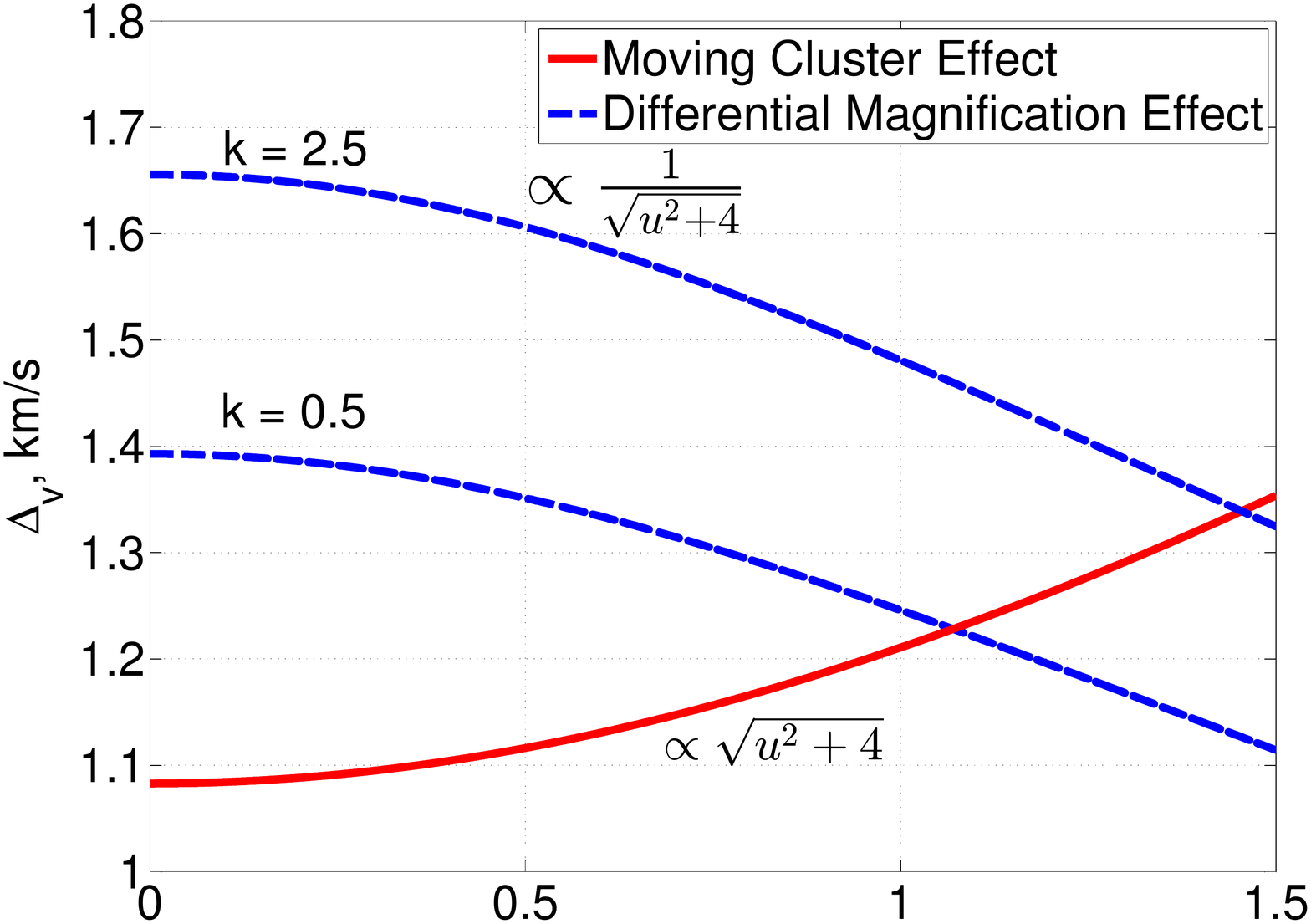}
	\caption{The difference in redshift between double images of a typical background galaxy as a function of its position, due to the effects described in the text (Equations \ref{MCE} and \ref{DME}). Parameter values used are listed in Table \ref{Inputs}. The shape of the rotation curve ($k$) has a modest impact on the DME once its flatline level $v_f$ is fixed. If instead $v_{max}$ is held fixed, then the impact of $k$ on the DME is very small.}
	\label{Overall_Effect}
\end{figure}

\begin{table} 
	\caption{Input parameters used for Figure \ref{Overall_Effect}. The source galaxy is assumed positioned so as to maximise the MCE (i.e. it is separated from the lens on the sky along the direction of motion of the lens, which is clear from images). The lens mass should roughly correspond to the sub-cluster in the Bullet. A flat $\Lambda CDM$ $\text{cosmology}$ is adopted \citep{Planck_Cosmological_Parameters}.}
	\label{Inputs}	
	%\centering
		\begin{tabular}{lll}
			\hline
			Parameter & Meaning & Value\\
			\hline
			$H_0$ & Present Hubble constant & 67.3 km/s/Mpc \\
			$\Omega_m$ & Present matter density & 0.315 \\
			$D_l$ & (Angular diameter)  & 0.945 Gpc \\
			& distance to lens at $z_l = 0.296$ & \\
			$D_s$ & Distance to source at $z_s = 1.7$ & 1.795 Gpc\\
			${D}_{ls}$ & Distance to source from lens & 1.341 Gpc\\
			& position in spacetime & \\
			$M$ & Mass of lens & $1.2 \times {10}^{14} M_\odot$\\
			$r_d$ & Scale length of source galaxy & 3.068 kpc\\
			$v_t$ & Tangential velocity of lens & 3000 km/s\\
			$v_f$ & Flatline level of source galaxy& 100 km/s\\
			& rotation curve & \\
			$\sin i \cos \gamma$ & See Figure \ref{Source_geometry} and Equation \ref{Angle_averaging} &$\frac{1}{2}$\\
			%Insert footnote indicating that actual value likely lower as edge-on disks are hard to spot.
			\hline
		\end{tabular}
\end{table}

The MCE $\propto v \sqrt{M}$ (Equation \ref{MCE}). Assuming also that $v \propto \frac{1}{M}$ for the components of the Bullet Cluster and that the value of $u$ would be broadly similar whichever component is targeted, the MCE overall $\propto \frac{1}{\sqrt{M}}$. This makes it larger around the sub-cluster. Furthermore, using its motion to extrapolate the total collision velocity is much more reliable than using the motion of the main cluster, because the sub-cluster contributes most of the relative velocity.

%extrapolating the collision velocity from the motion of the main cluster is much less reliable than using the sub-cluster (which contributes most of the relative velocity).

A typical source galaxy orientation is chosen using Equation \ref{Angle_averaging}. Although we used $v_f = $ 100 km/s, it is around double this for our own galaxy \citep[e.g.][]{McMillan_2011}. 

Using the parameter values in Table \ref{Inputs}, we obtained the results shown in Figure \ref{Overall_Effect}. The DME and MCE are comparable in magnitude. 

Observing a similar source multiply imaged by the higher mass component instead does not reduce the relative importance of the DME. This is because the DME $\propto \frac{1}{\sqrt{M}}$ (Equation \ref{DME}), just like the MCE. In fact, this scaling highlights an additional problem: substructure in the lens (e.g. individual galaxies) with much less mass than the entire cluster can enhance the DME. For example, an elliptical galaxy in the lens plane with $M = 10^{13} M_\odot$ would cause a DME $\sim$3 times larger than the smooth cluster potential.

This problem could be mitigated to some extent by not selecting images which show indications of being lensed by small scale structure (e.g. avoiding images appearing near a galaxy in the lens plane). We have implicitly assumed that such a selection has been done, such that a point mass model for the lens is appropriate. Even in this case, it might well be necessary to correct for the $\text{Differential}$ $\text{Magnification}$ $\text{Effect}$. This correction would be less relevant if targets could be selected for which the Effect is small. We now consider how these things might be achieved.

%\newpage
%\clearpage
\section{Correcting For The Effect}
\label{Adaptation}

%Ultimately, obtaining better than 1 km/s spectral resolution in a multi-pixel image of the source galaxy would allow one to calculate the difference in redshift between the same part of the galaxy in the two images. This would essentially eliminate the DME, because the magnification would be nearly uniform over a small part of the galaxy and because the radial velocity hardly varies over it. However, the wait for such high quality data may be long.

For spatially unresolved spectra, it is possible to calculate the DME by determining the parameters in Equation \ref{DME}. If radial velocities accurate to a few km/s are obtained for a galaxy, then determining ${v}_{max} \sin i$ to $\sim$10 km/s should be feasible using widths of spectral lines (see later). 

$r_d$ might be obtained from an image of the target, once distortion and magnification by the lens were corrected for. If the image were taken at more than one wavelength, it would suggest a value for $k$ (which we don't need very accurately) as the colour can be used to estimate the baryonic $M$/$L$.

There is no need to determine the inclination as we are only interested in redshift gradients across the source. However, the orientation of the major axis of the image is important in determining the axis of rotation and thus the angle $\gamma$ in Figure \ref{Source_geometry}. 

To know the sense of rotation (i.e. which side of the source galaxy is the approaching side), we would need spectra of different parts of the source galaxy. Naturally, a disk-integrated spectrum would be insufficient for this purpose. However, one could make do with poorer spectral resolution. 

The secondary image is inverted relative to the primary, providing an important consistency check if both images were used for such a determination. We strongly recommend doing this, because an error would lead to a 200\% error in the calculation of the DME. The chance of this is minimised with two determinations of the sense of rotation. 

Finally, we also need $\bmath{\nabla} A$, which must come from a lensing reconstruction.

%Also required is the sense in which the source galaxy rotates, i.e. which side is the approaching side. This can't come from a disk-integrated spectrum or an image, and requires a spectrum for different sides of the source galaxy.

\subsection{Additional Information From Detailed Spectral Line Profiles}
\label{Detailed_Line_Profiles}

\begin{figure}
	\centering
		\includegraphics [width = 7cm] {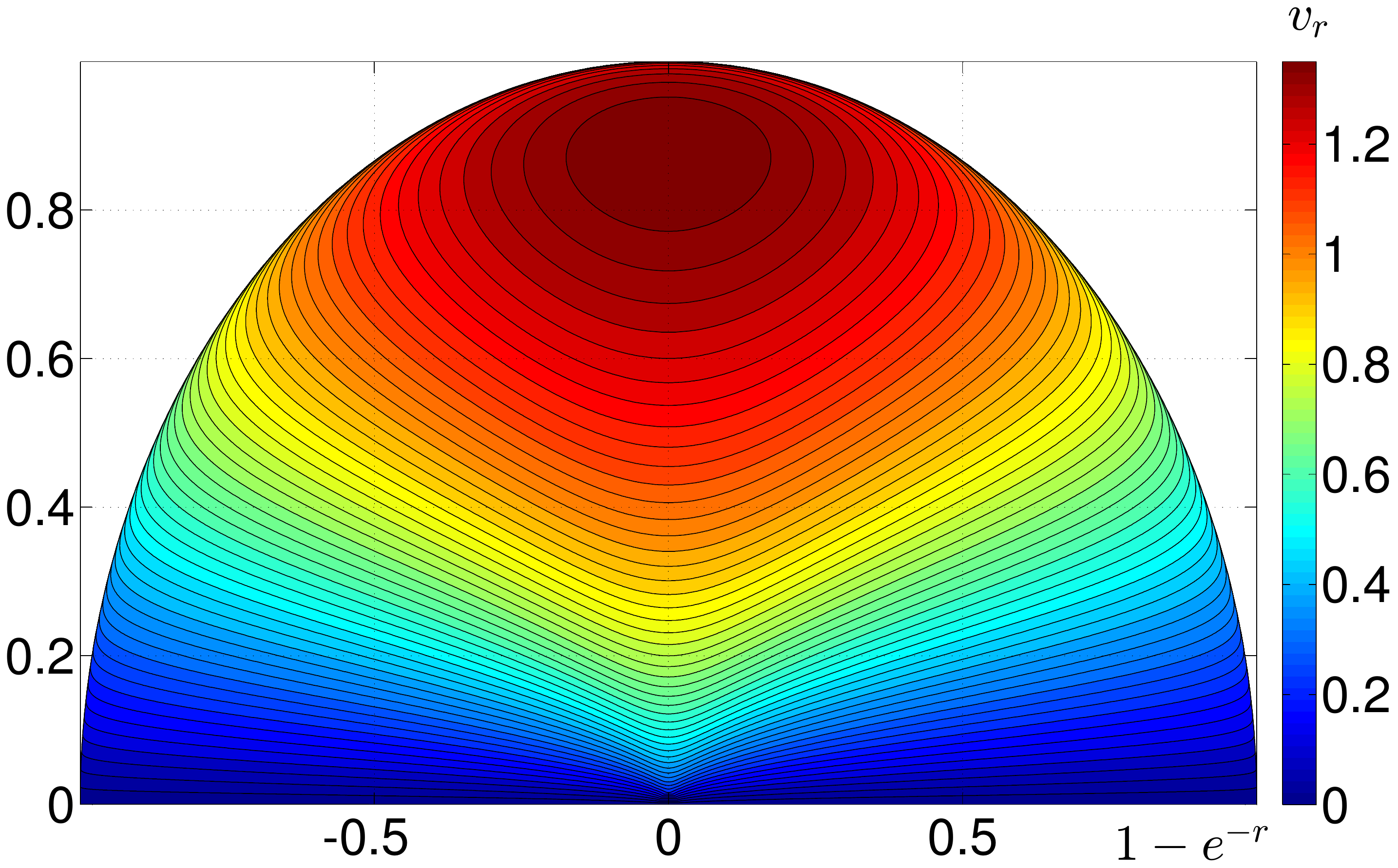}
	\caption{Radial velocity map of a disk galaxy viewed by an observer within its plane at large $x$, for the case $k = 2.5$. Radial velocities are antisymmetric about the $x$-axis. The radial co-ordinate is rescaled so all parts of the figure would be equally bright. The units are such that $r_d = 1$ and $v_f = 1$. Note the large region with $v_r$ close to the maximum value. The result for $k = 0.5$ is very similar, although $v_{max}$ is much closer to $v_f$.}
	\label{Radial_Velocity_Contour_k_2_5}
\end{figure}

\begin{figure}
	\centering
		\includegraphics [width = 8.5cm] {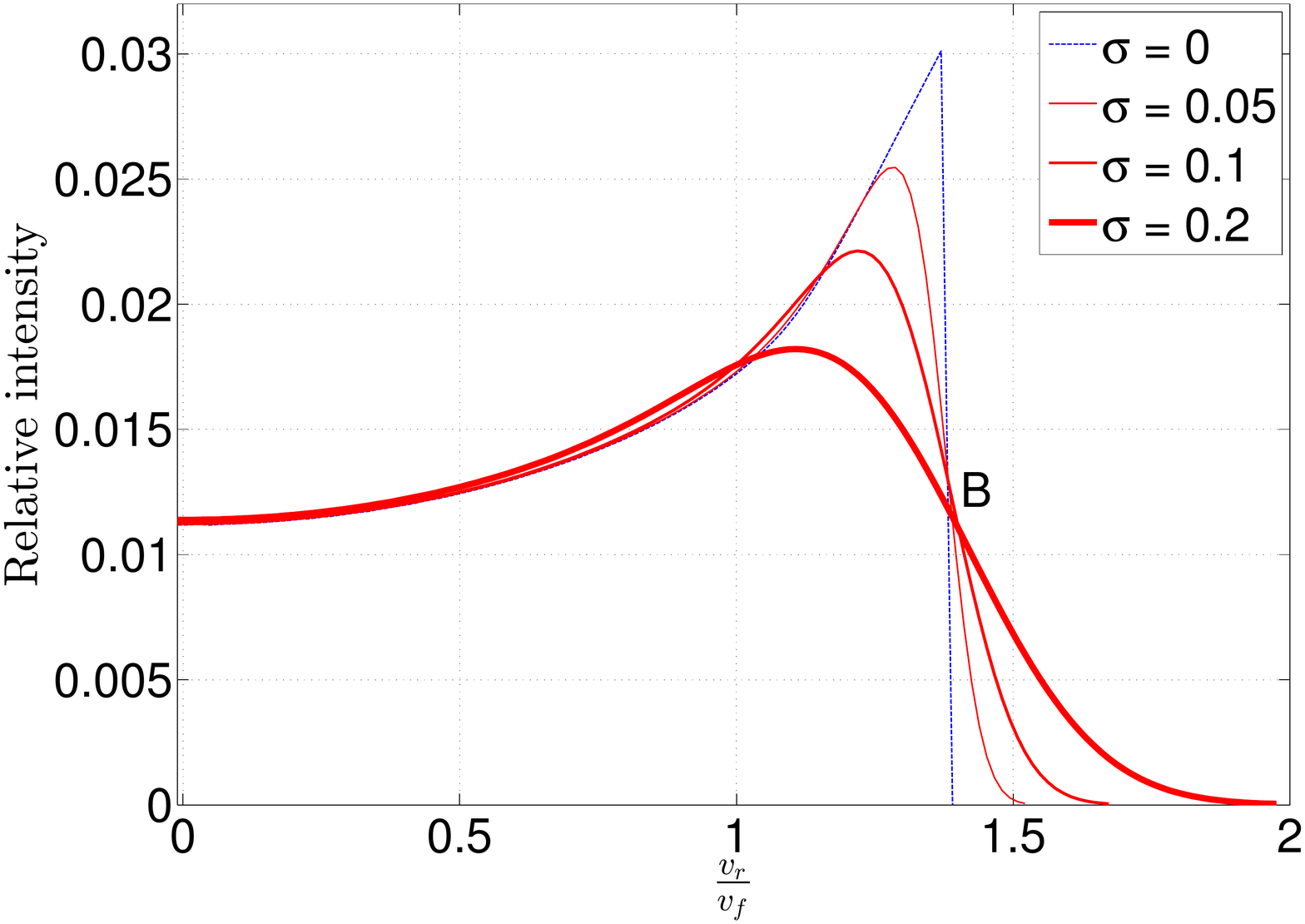}
	\caption{The synthetic line profile of an intrinsically narrow line in an unlensed galaxy with $k = 2.5$, viewed edge-on. The profile is symmetric about $v_r = 0$. Velocities are scaled to the flatline level $v_f$. The sharp drop in the line profile (blue) would probably get blurred (e.g. by random motions), so we convolved the profile with Gaussians of widths $\sigma$. The results are shown as red lines with thicknesses $\propto \sigma$. Notice how all 4 profiles pass close to the point marked $B$. The result for $k = 0.5$ is similar, if the profiles are scaled to have the same $v_{max}$ rather than $v_f$.}
	\label{Spectral_profiles}
 \end{figure}

It is often possible to extract more information from a spectral line than just the location of its centroid. The width of the line profile can be used to estimate e.g. $v_{max} \sin i$.

The MCE simply shifts the entire spectrum. The DME leads to a `tilt' being introduced because one side of the galaxy is magnified more than the other. These effects are different. Therefore, detailed line profiles can tell us if the shift in the centroid of spectral lines is due to the MCE or the DME. This would avoid the need to determine parameters like the disk scale length and orientation. A detailed lensing reconstruction to determine $\bmath{\nabla} A$ would also be avoided. 

We investigate how the DME and MCE affect line profiles of disk galaxies with rotation curves parametrised by Equation \ref{Rotation_Curve}. We assume the galaxy is viewed edge-on, so the radial velocity of any part of it is
\begin{eqnarray}
	v_r(r, \phi) = v_c(r) \sin(\phi)
\end{eqnarray}

The resulting radial velocity map is shown in Figure \ref{Radial_Velocity_Contour_k_2_5}. Only half of the galaxy is shown because $v_r$ is antisymmetric about the viewing direction (the $x$-axis). $v_r$ is symmetric about the $y$-axis, because $\sin \phi = \sin \left( \pi - \phi \right)$.

To determine the profile of a narrow spectral line, we divide the galaxy up into a large number of elements. We use cylindrical polars so $v_r$ becomes separable. Thus, the rotation speed only needs to be calculated once at each $r$ (for all $\phi$). The radial velocity at the centre of each element is used to classify it among 200 bins in radial velocity.

Assuming constant M/L for the baryons, the total intensity of the element multiplied by the magnification $A$ is then assigned to the corresponding radial velocity bin. Because radial velocities and wavelengths are directly related, in this way one obtains a synthetic line profile. 

%\footnote{The radial velocity is separable in cylindrical polars, so the rotation speed only needs to be calculated once for all $\phi$ at a given $r$.}

Spectral lines have an intrinsic width and can be further broadened by random motions within the galaxy. To account for these effects and also for instrumental errors, we convolved our synthetic line profiles with Gaussians of various widths $\sigma$. The results are shown in Figure \ref{Spectral_profiles}. The sharp peaks at $v_r \approx \pm {v}_{max} \sin i$ give rise to the name of a double-horned profile. 

These horns are caused by the rotation curve having a peak, leading to a small range in $v_r$ corresponding to a large range in $r$. The greatest attained values of $\left| v_r \right|$ also correspond to large ranges in $\phi$, because $\sin \phi$ is nearly independent of $\phi$ when $\phi \approx \pm \frac{\pi}{2}$. 

Thus, a small range in $v_r$ corresponds to a large region in the galaxy. Moreover, the peak rotation speed occurs at a radius close to that which maximises the light emitted per unit radius ($r = r_d$). Figure \ref{Radial_Velocity_Contour_k_2_5} shows the `bull's-eye' corresponding to the fairly large region with near-maximal $\left| v_r \right|$. This is responsible for the very pronounced horns in the line profile. They are somewhat less pronounced at high $\sigma$. 

\begin{figure}
	\centering
		\includegraphics [width = 8.5cm] {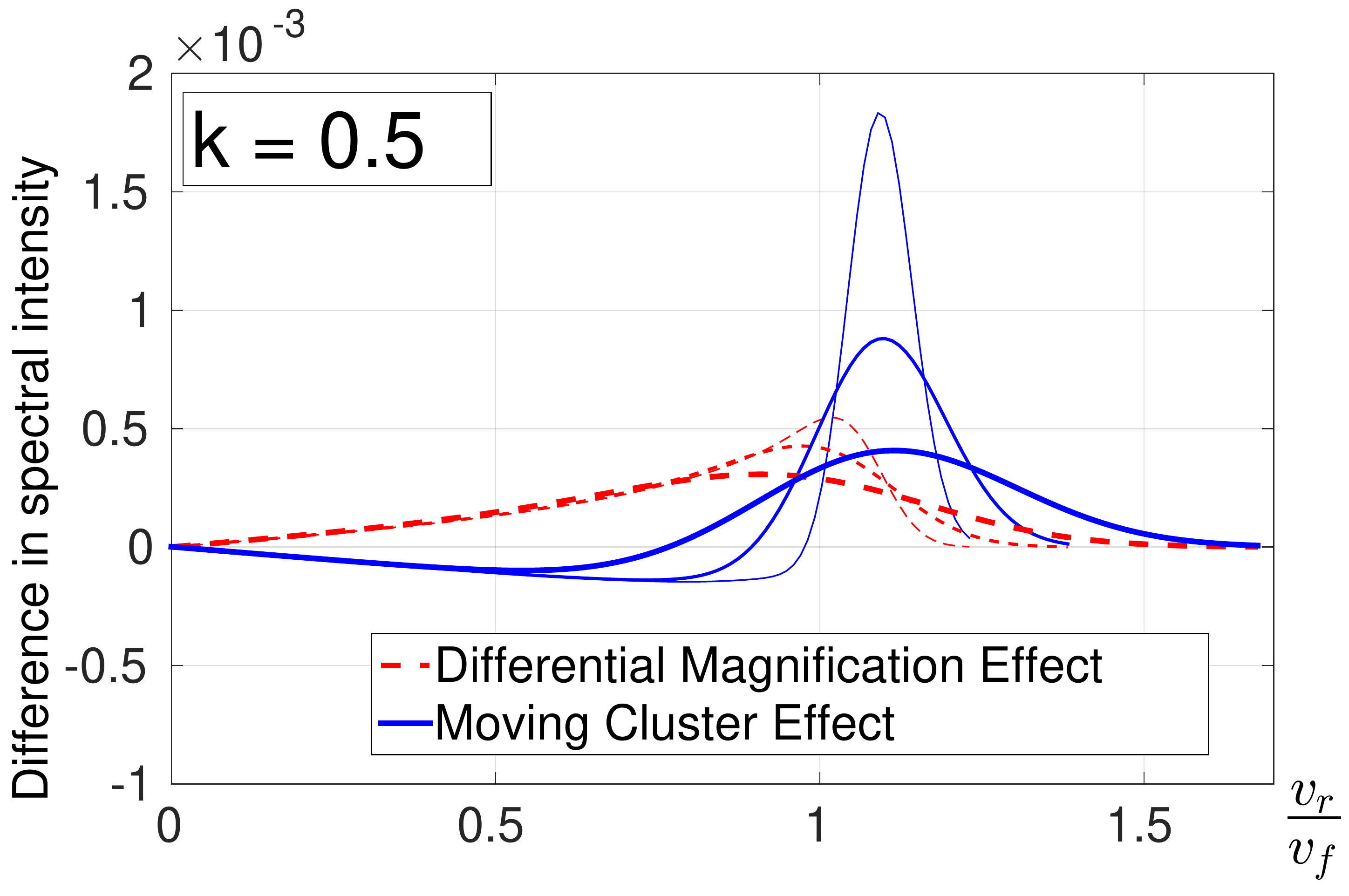}
		\includegraphics [width = 8.5cm] {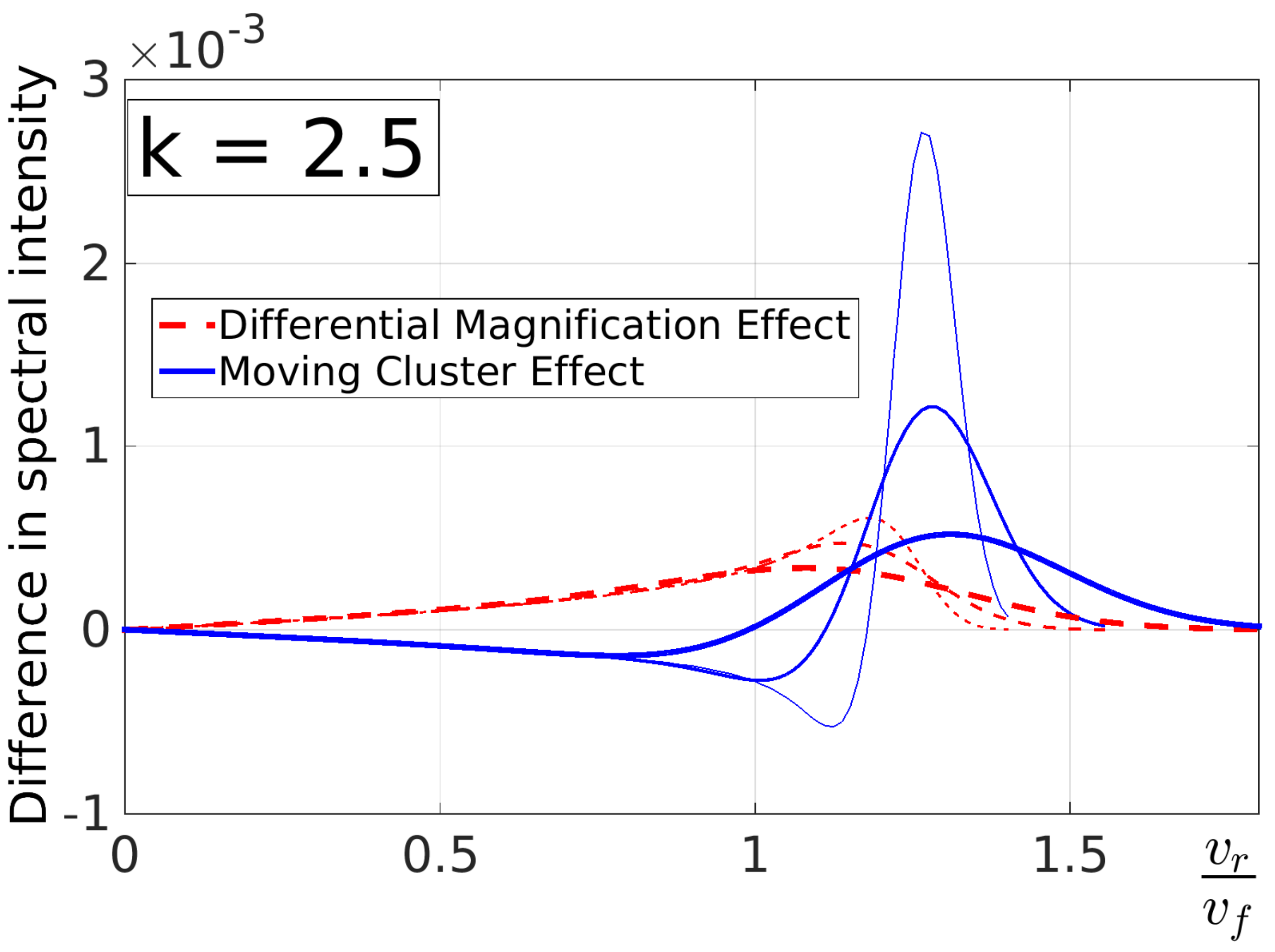}		
	\caption{The residuals in the spectral profile due to the DME (Equation \ref{Equation_39}) and the MCE (horizontal shift of profile) are shown here. These were obtained by subtracting a control line profile (Equation \ref{Control}). The patterns are antisymmetric about $v_r = 0$. Results are for an edge-on galaxy with $k = 0.5$ (top) and $k = 2.5$ (bottom). Both effects change the mean redshift by 1\% of the maximum rotation speed, representing 1.08\% of $v_f$ for $k = 0.5$ and 1.26\% for $k = 2.5$. The spectra were convolved with Gaussians of widths 0.05, 0.1 and $0.2~v_f $ (higher $\sigma$ indicated by thicker line). The MCE can't change the amplitudes of the horns. The DME makes one more pronounced and the other less.}
	\label{Residuals_k_0_5}
\end{figure}

Although one might expect a feature corresponding to $v_f$ (at least at low $\sigma$), this is absent. A quick look at Figure \ref{Radial_Velocity_Contour_k_2_5} shows why: $v_c \left( r \right) \approx v_f$ only for sufficiently large $r$. There is very little light from such regions, so a disk-integrated spectrum is hardly sensitive to them. In fact, due to the steep decline in surface brightness with $r$, most of the spectral intensity at $v_r = v_f$ actually comes from the rising part of the rotation curve (when $v_c \sin \phi = v_f$) rather than from the flat part. Thus, in the line profile, there is nothing special about $v_f$. This is not true for $v_{max}$.

%One might expect a feature at $v_r = v_f \sin i$. Figure \ref{Radial_Velocity_Contour_k_2_5} shows why this is absent: the rotation curve converges to $v_f$ only at large radii, where is little baryonic mass (and thus light) so far out. However, the `bull's eye' corresponding to $v_r \aprox v_{max} \sin i$ is clearly apparent. This is responsible for the horns in the line profile.

%For the case $\sigma = 0$, no feature is apparent at $v_f$. This is because the rotation curve is close to $v_f$ only at large radii. Such regions are not very bright and so contribute little to the spectrum. 

Determining $v_{max} \sin i$ from a line profile is non-trivial as the horns move to lower $\left| v_r \right|$ as $\sigma$ increases. Instead of using the horn positions, one could use the values of $v_r$ where the intensity is a certain fraction of the intensity at the line centre ($v_r = 0$). If this fraction is chosen carefully, then one could simply scale the resulting $v_r$ by a constant factor and accurately recover $v_{max} \sin i$ over a wide range in $\sigma$ and $k$. To see why, note that spectra with different $\sigma$ all pass close to the point marked $B$ in Figure \ref{Spectral_profiles}. 

%$k$ has only a modest impact on the spectral profile if plotted against $\frac{v_r}{v_{max} \sin i}$, while $\sigma$ mostly affects the tails. 

Ultimately, it might be better to compare the observed line profile with a suite of synthetic profiles built for a range of $k$, $\sigma$ and $v_{max} \sin i$. The initial guess for $\sigma$ might come from considering the shape of the tail. If $v_{max} \sin i$ is accurately recovered, then the DME hardly depends on $k$. \footnote{Line profiles can also be used to find $v_f \sin i$ without detailed rotation curves. In this case, the value of $k$ is important as a `$\Sigma-$correction' must be applied to get from $v_{max}$ to $v_f$ (e.g. bottom panel in Figure \ref{Model_Rotation_Curve}). $k$ doesn't affect the line profile much and so it would need to come from an image \& photometry.}

%In principle, the rotation curve flatlines at $v_f$ and this should also lead to a sharp spectral feature at $v_f$. This is absent because the rotation curve flatlines very gradually. Thus, by the time it falls close to $v_f$ again, the radius is very large and the surface density is very small. This means that, to estimate $v_f$ from spectral line profiles, one should apply a correction factor similar to that in Figure \ref{Model_Rotation_Curve} to the positions of the horns (as well as an inclination correction). However, it is possible that with a different rotation curve shape, there might be a feature at $v_r = v_f$. In any case, $v_f = v_{max}$ for LSB galaxies (i.e. low $k$).

%We have assumed that the line has no intrinsic width and the galaxy has no velocity dispersion. Thus, spectral features in Figure \ref{Spectral_profiles} will be somewhat less sharp in real data. However, if the spectral line arises from gas, then the velocity dispersion is likely to be only a small fraction of $v_f$, making the effect not very important.

%Synthetic spectral profiles corresponding to very narrow lines are shown, for two choices of rotation curve shape ($k$) and a galaxy viewed edge-on. The spectrum is symmetric about $v_r = 0$ so only half of it is shown. For each $k$, the spectra were convolved with three different velocity dispersions, with thicker lines indicating higher $\sigma$. The values used are $\frac{\sigma}{v_f} = $ 0.05, 0.1 and 0.2. The horn at $v_r = {v}_{max}$ is clearly apparent.

The horns are caused by a relatively small part of the galaxy but they greatly affect the mean radial velocity of its image. Thus, if the galaxy was not axisymmetric and e.g. had a dusty spiral arm obscuring light from this region, then the redshift measurement of each image may be biased. Partly for this reason, it may be a good idea to consider the rest of the line profile and not just the mean redshift (which is basically the same as considering just the horns).

%This region is close to the radius where the rotation curve peaks and close to $\phi = \frac{\pi}{2}$. 

We now consider how the DME and MCE affect the line profile. The mean redshift velocity of the line is raised by 1\% of $v_{max}$ (1.08\% of $v_f$ for $k = 0.5$ and 1.26\% for $k = 2.5$). We consider separately the cases where either the DME or the MCE is wholly responsible for this shift in line centroid. We also construct control line profiles like those in Figure \ref{Spectral_profiles}. These are obtained by setting 
\begin{eqnarray}
	A = 1 ~ {\forall}_{r,\phi}
	\label{Control}
\end{eqnarray}	

In Figure \ref{Residuals_k_0_5}, we show the pattern of residuals (relative to the control) created by each effect. The total line intensity is kept the same for the comparison.

To obtain the corresponding observations, one would need to account for the images having different overall magnifications. Thus, the spectra would have to be rescaled. We assume this can be done perfectly (i.e. the photometry is very accurate). 

%We show the pattern of residuals between the same spectral line in two images of a background galaxy caused by the DME and the MCE. We assume the total intensities of the images have been matched. The magnifications of the images are usually different, so one of the images has to be artificially brightened relative to the other. We assume this matching can be done perfectly (i.e. the photometry is very accurate). 

%The mean redshift difference between the images is fixed at 1\% of the maximum rotation velocity (1.08\% of $v_f$ for $k = 0.5$ and 1.26\% for $k = 2.5$). The pattern of residuals is shown for the case where either the DME or the MCE is wholly responsible for the difference in mean image redshift compared to a control spectrum. Such a spectrum is obtained simply by setting $A = 1$.

The MCE corresponds to a horizontal shift in the spectrum relative to the control. This means the amplitudes of the horns are unaffected. \emph{The pattern of residuals corresponds to the gradient in the spectrum}. Thus, the residuals are largest near the positions of the horns, but small precisely at them. The residuals are of opposite signs on either side of each horn because the gradient in the spectrum changes sign there.

For the DME, we set 
\begin{eqnarray}
	A = 1 + n \widetilde{r} \sin \phi
	\label{Equation_39}
\end{eqnarray}	

Note that no assumptions are made about any of the factors controlling the amplitude of the DME, beyond it being a small effect relative to $v_{max} \sin i$ (i.e. $n \ll 1$) and that we need not consider the second-order dependence of $A$ on position in the source plane. The purpose here is to illustrate how the DME affects the line profile, not how much (this is controlled by $n$). If the DME $\sim$$0.01 v_f \sin i$, then the residuals would be $\sim$1\% of the line profile.

The image overall is not magnified for any (small) $n$. We adjust $n$ until the line centroid shifts by the correct amount, to allow comparison with the MCE.

%A velocity dispersion of $0.1 v_f \sin i$ is also included in the calculations (i.e. the spectra were convolved with a Gaussian of this width before determining differences between them).

The DME causes one side of the galaxy to be magnified more than the other. Thus, the residuals due to it are of the same sign for each half of the galaxy (e.g. for $v_r > 0$). There is no change in sign at the horns.

These correspond to material displaced from the centre of the galaxy along the direction $\bmath{OP}$ in Figure \ref{Source_geometry}. As argued previously, we only need to consider the component of $\bmath{\nabla} A$ along this direction. Thus, the effects of differential magnification are substantial for the material corresponding to the horns (in so far as the DME affects the image at all). This is in contrast to the MCE, which hardly affects the line profile at these positions (because the gradient of the line profile there is 0).

%This means the DME has maximal impact on the line profile close to $v_r = \pm {v}_{max} \sin i$. 

For some $v_r$, the MCE leads to very large residuals if $\frac{\sigma}{v_f \sin i}$ is low (Figure \ref{Residuals_k_0_5}). Thus, observing faster-rotating galaxies might make it easier to distinguish between the DME and the MCE (as $\frac{\sigma}{v_f \sin i}$ would likely be smaller). However, the DME would be larger and so it would have to be accounted for more accurately.

Detailed profiles of individual spectral lines may therefore help in determining the balance between the MCE and the DME in accounting for redshift differences between multiple images. In reality, a large number of spectral lines would probably need to be stacked. Even then, it seems likely that, in so far as redshift differences between the images are discernible, the cause of such differences can also be determined. 

\subsection{The Second Order Effect}

%For the case considered, our calculation of the DME is hardly affected by the second-order dependence of $A$ on sky position. However, this might still affect the spectra more substantially, albeit in a symmetric way.

%In general, $A_0$ ($\equiv A$ at the centre of the source galaxy) isn't equal for the images but spatial derivatives of $A$ are equal for them. Thus, the fractional change to the total intensity of each image will be unequal. This means that the DME calculated so far can't simply be multiplied by some correction factor to account for second order effects, as the correction will depend on $u$.

%The second-order dependence of $A$ on sky position affects the total intensity of each image. $\frac{\partial^2 A}{\partial u^2} > 0$ so in Figure \ref{Source_geometry}, points in the galaxy separated from the centre in the direction of $\bmath{OP}$ are magnified more than $O$. As $A$ decreases with $u$ and points separated along $\bmath{OQ}$ have greater $u$ at second order, such points are magnified less. These effects therefore partially cancel.

For simplicity, we continue assuming the source is located along $\bmath{OP}$. Thus, regions with high $\left| v_r \right|$ are magnified more and regions with lower $\left| v_r \right|$ are magnified less than the centre of the source due to the second order dependence of $A$ on position. To investigate what this means for spectral line profiles, we set $A$ to depend quadratically on position along the direction $\bmath{OP}$. This introduces a $\sin^2 \phi$ dependence. 
\begin{eqnarray}
	A = \frac{1 + n \widetilde{r}^2 \sin^2 \phi}{1 + 3n}
	\label{Quadratic}
\end{eqnarray}

A quadratic dependence along the orthogonal direction would give a $\cos^2 \phi$ dependence. Because $\cos^2 \phi = 1 - \sin^2 \phi$, a second derivative of $A$ in either direction would affect the line profile in the same way (i.e. the residuals would have the same pattern, up to sign); once any overall magnification was corrected for.

When comparing the spectra, observers would first scale them to have equal intensities. Thus, we must avoid changing the total intensity. This leads to the factor of $1 + 3n$ in the denominator of Equation \ref{Quadratic}.  

\begin{figure}
	\centering
		\includegraphics [width = 8.5cm] {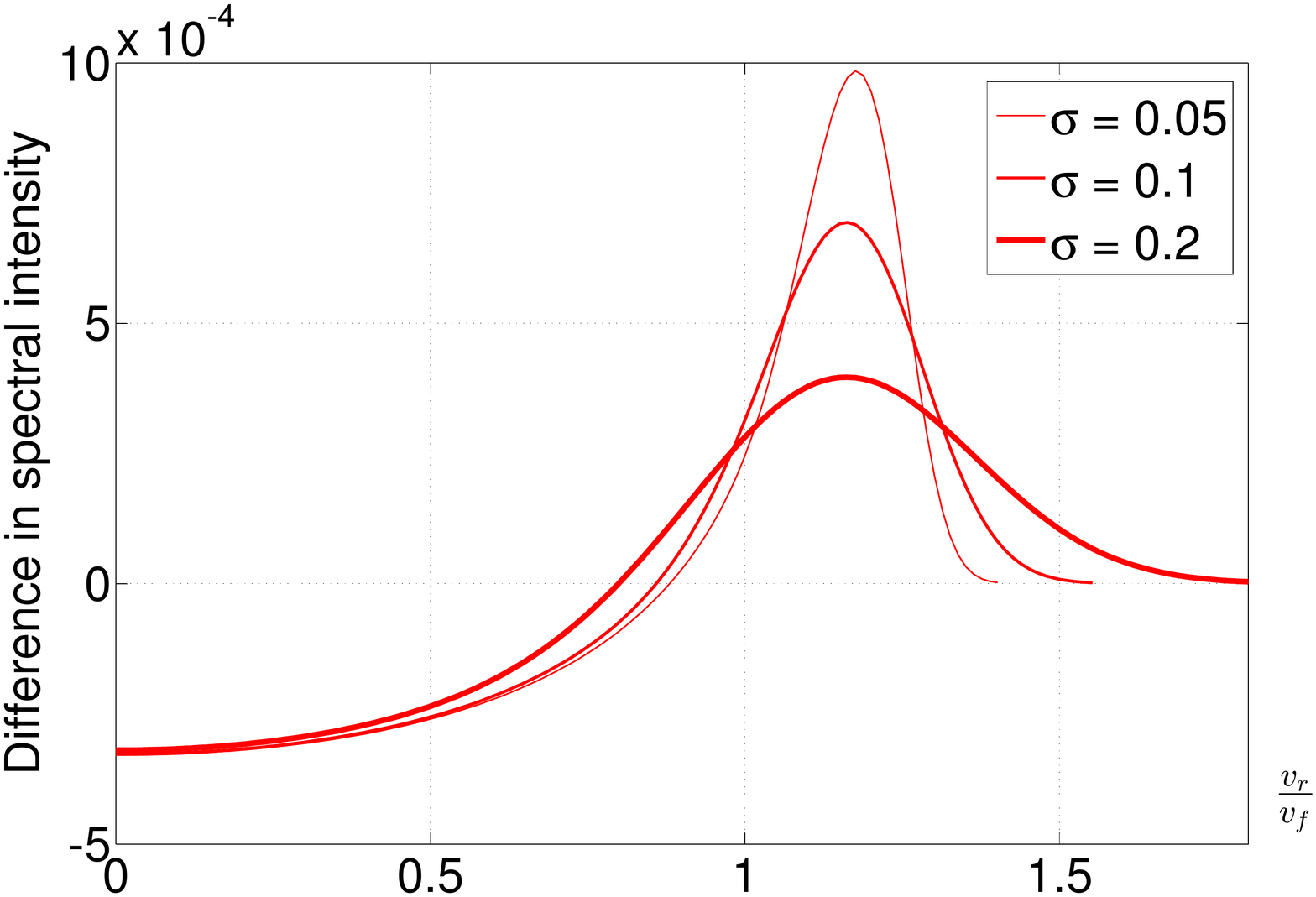}
	\caption{The pattern of residuals for the second-order DME (Equation \ref{Quadratic}) and $k = 1$. A control profile obtained using Equation \ref{Control} was subtracted and the result convolved with a Gaussian. The residuals are symmetric about $v_r = 0$, so both horns become more pronounced in this example.}
	\label{Quadratic_k_2_5}
\end{figure}

The results are shown in Figure \ref{Quadratic_k_2_5}. The effect is symmetric in $v_r$, so both horns are equally affected. These end up more pronounced in the secondary image than in the primary (in this example).

In reality, both the first and second order DME would be present for any given pair of images of the same object. Thus, the residuals would have both an antisymmetric  and a symmetric part. However, the latter would likely be very small for cluster mass lenses (except for caustic images).

%A more accurate calculation of $\overline{v_r}$ for each image including this second order effect is beyond the scope of this work. Such effects might be important for galaxy-galaxy lensing as $\theta_E$ is smaller.

\subsubsection{Non-rotating sources}

We briefly consider how the DME might affect a non-rotating pressure-supported galaxy, such as an elliptical. If a galaxy is symmetric such that $\rho(\bmath{r}) = \rho(-\bmath{r})$ and $\sigma(\bmath{r}) = \sigma(-\bmath{r})$, then at first order the DME does not affect an unresolved image at all. To see this, consider an inversion mapping $\bmath{r} \to -\bmath{r}$ while leaving $\bmath{\nabla} A$ unchanged. The situation is identical to reversing the direction of $\bmath{\nabla} A$ instead, so one expects the DME to act in exactly the opposite way on the spectrum. But the situation has not physically changed, so the DME must also remain unchanged. 

This conclusion breaks down at second order. Suppose parts of the galaxy further from its centre are magnified more. Then, as the velocity dispersion generally decreases outwards, the derived velocity dispersion of the image will be reduced by the DME. The effect is larger for the fainter (secondary) image, which will thus appear to have a smaller velocity dispersion than the primary in this example.

This is likely to be more important for galaxy-galaxy lensing as $\theta_E$ is smaller, making $du$ over the source larger. In this case, the DME might be useful to constrain the form of $\sigma(r)$ using a double image of a distant elliptical galaxy. 

Alternatively, if the source galaxy was well-understood, one might be able to constrain $\bmath{\nabla} A$ and thus have a better understanding of the lens. Doing both simultaneously would likely be very challenging and model-dependent.

%\newpage
%\clearpage
\section{Targets With A Smaller Effect}
\label{Mitigation}

Figure \ref{Overall_Effect} shows that the Differential Magnification Effect may well need to be accounted for when using the redshifts of double images to determine the motion of the lens. However, doing this accurately may be difficult because of the cosmological distance to the source galaxy. Therefore, we suggest sources for which the DME should be smaller, allowing us to correct for it less accurately. 

Some strategies outlined here involve selecting targets which are harder to observe, thereby making their spectra less accurate. It is up to observers to decide which targets best minimise the uncertainty introduced by the DME while still being feasible to obtain accurate spectra for. We also note that minimising the uncertainty introduced by correcting for the DME is not necessarily equivalent to minimising the magnitude of the DME, because there may be sources for which the DME can be estimated more reliably.

\subsection{QSOs}

The DME $\propto r v$, where the source has typical size $r$ and radial velocity spread $v$. For a given mass $M$, the Virial Theorem gives $v \propto \frac{1}{\sqrt{r}}$. Thus, the DME $\propto \sqrt{r}$. For sources with a particular $M$, the DME would be reduced if the source were smaller, even though it would spin faster. 

One obvious type of very small target visible over cosmological distances is a quasi-stellar object (QSO).  If a doubly imaged QSO could be found lensed by the Bullet Cluster, it might make an excellent target.

QSO spectra can sometimes lack distinctive features which are required for precise redshift measurements. The Ly-$\alpha$ forest might provide a solution, but only if the same feature appeared in both images. Because the rays of light corresponding to the images diverge from the source\footnote{by an angle $\frac{1 + z_l}{1 + z_s} \frac{D_l}{D_{ls}} \left( \theta_1 - \theta_2 \right)$}, this is only feasible if the gas cloud causing the absorption feature was located fairly close to the QSO. 

Another problem might be that the small size of QSOs makes their radiation time-variable. Thus, the time delay between the images could make it difficult to compare their spectra. This might require observers to wait out the time delay, which would first have to be determined (though it could be estimated, perhaps using Equation \ref{Time_delay}).

\subsection{Smaller \& Fainter Galaxies}

The DME is proportional to both the rotation velocity and the size of the source galaxy. Brighter galaxies generally rotate faster \citep{Tully_Fisher_1977}, so targeting fainter galaxies might help. One advantage of this approach is that the number density of fainter galaxies is greater than for brighter ones \citep{Schechter_1976}. This makes it more likely that a suitably oriented multiple image can be found.

However, it would be harder to obtain accurate spectra $-$ and thus redshifts $-$ for fainter targets. Given the high accuracy required in the redshift measurements and the cosmological distance to the source, this is perhaps not the best option at present.

\subsection{Elliptical Galaxies}

%All parts of such galaxies have very little/no net radial velocity, making the DME much less important. Elliptical galaxies may be distinguished by their shape (and possibly colour). If an image is highly flattened, then it is not an elliptical. But if it is circular, then it is irrelevant whether it is an elliptical or a face-on spiral as neither will have systemic motions along the line of sight so neither suffers from a DME. Thus, we only need be concerned about flattened-looking galaxies.

%There is another key difference between spiral and elliptical galaxies if one has a very high resolution spectrum, as would be required to measure the MCE. With a spatially unresolved spectrum, spiral galaxies typically have a double-horned profile for any spectral lines (see next section). 

Elliptical galaxies might make good targets as they usually rotate slower than spirals, if at all. They might be distinguished using colour or image shape (though one might need to correct for distortion by the lens). The surface brightness declines outwards much more gradually for ellipticals than for spirals, potentially providing another way of finding them. 

Before conducting detailed observations, targets selected like this might be followed up to check if the spectral line profiles were double-horned (characteristic of rotation along the line of sight). A good target should have a Gaussian-looking line profile, characteristic of a pressure-supported object.

%If the target was a face-on spiral, then the dispersion would be very small. Crucially, there would be no redshift gradient across the image in this case.

However, even ellipticals can rotate, so the DME might not be eliminated by observing one. Also, most galaxies are not elliptical, so finding a bright doubly-imaged one is somewhat dependent on luck. Nonetheless, we consider this the best option. This is partly because the work of \citet{Gonzalez_2009} identified a multiply imaged galaxy which may be a good target for determining the MCE.

\subsection{Galaxy Orientation \& Viewing Angle}

Figure \ref{Source_geometry} shows the geometry of the situation. The radial velocity of any part of the galaxy is scaled by $\sin i$, so a face-on spiral could not have a redshift gradient across it and thus would be unaffected by the DME. 

Determining $i$ requires an image of the source galaxy to determine its shape. One could imagine trying to select targets which look round. Even then, the target might be an elliptical galaxy with some rotation along the line of sight.

The direction of $\bmath{\nabla} A$ is also very important. In theory, we should seek situations where $\bmath{\nabla} A$ is orthogonal to the major axis of the image. In such situations, an edge-on disk galaxy would appear as a line on the sky aligned orthogonally to the direction towards the lens. With more complicated lenses, the galaxy would appear as a line on the sky orthogonal to $\bmath{\nabla} A$, which hopefully could be estimated using a lensing reconstruction.

\subsection{Galaxy Position \& Caustics}

As was already pointed out in \citet{Molnar_2013}, the MCE is maximal for image separations aligned with the direction of the lens' proper motion. As the collision is nearly in the plane of the sky, the likely direction of this motion is plain to see \& observers should target double images separated approximately along this direction.

In the simple lens model that we use, the DME $\propto \frac{1}{\sqrt{u^2 + 4}}$ while the MCE $\propto \sqrt{u^2 + 4}$. Thus, galaxies less closely aligned with the lens make better targets in terms of the systematic error of the DME. For such sources, the images are more widely separated.

Unfortunately, sources with larger $u$ make worse targets under a number of other considerations. Both images $-$ but especially the secondary $-$ are fainter. This image also becomes very close to the lens, making it more likely to be obscured.

A lensing reconstruction could be used to suggest particular locations where the magnification is nearly constant. A galaxy with double images near such locations might make a good target for measuring the MCE. The difficulty with this is that such `sweet spots' might be small and not have any observable galaxies in them. Also, a magnification map of sufficient accuracy might be difficult to obtain. 

Regions where $A$ varies rapidly with position enhance the DME. Caustics occur where the magnification of a small part of the source plane is infinity. This means that the magnification varies rapidly with position in the source plane, greatly increasing the DME. For this reason, it has been suggested to avoid caustic images \citep{Molnar_2013}.

However, it may be worthwhile to try correcting for the DME in caustic images because they are generally very bright, making for more accurate spectra. The correction would need to be done very accurately in this case, because the MCE might be much smaller than the DME. Because this is likely to lead to controversy surrounding the measurements, we also recommend avoiding caustic images unless the observational case is compelling.

\subsection{Substructure Within The Source Galaxy}

If the source galaxy has e.g. a bright star-forming region which emits strongly in the UV while the rest of the galaxy does not, then another possibility arises. UV spectral lines would correspond to material in a small part of the galaxy. Consequently, different parts of it would have much the same radial velocity and the magnification across it would be more uniform than across the whole galaxy. This would reduce the DME for the UV lines.

Thus, in this example, the redshifts for the images should be calculated using only the UV lines. In practice, one would exploit the fact that a small part of the galaxy should have only a narrow range of redshifts. Thus, one might use only spectral lines which have a similar redshift. If the intrinsic linewidth was small, then the line should be very narrow as there would not be much rotational broadening.

Another possibility is using spectral lines that are more prominent in the bulge of the galaxy (if it has one). The bulge is mostly pressure-supported with little rotation and is also much smaller than the whole galaxy. In this case, the spectral lines to use might be quite broad, but have a Gaussian line profile even if the galaxy is rotating (so most spectral lines have a double-horned profile). 

For this technique to be of much benefit, the galaxy needs to be quite inhomogeneous in some way. It might be difficult to tell whether it is from an image. Also, the technique reduces the number of spectral lines that are used to calculate the redshift, making for less precise measurements. This makes it difficult to target fainter galaxies, perhaps forcing observers to choose between observing all of a fainter galaxy or (effectively) part of a brighter one.

Any decision to restrict which spectral lines are used to determine the MCE should be justified based on more detailed observations of nearby galaxies. This increases confidence that the decision does indeed effectively restrict the observations to a small part of the source. 

Even if all usable spectral lines are used to measure the MCE, it is still likely that some lines are less affected by the DME than others. It may be important to allow for this in the analysis e.g. by grouping spectral lines based on their linewidth and shape and obtaining an inference on the MCE for each group. 

Due to the uncertainties introduced by such procedures, we recommend reducing the DME by careful choice of target so that the exact method used to correct for it does not much affect the inferred lens velocity.

\section{Observational prospects}
\label{Observations}

We now consider the technical feasibility of detecting the MCE with high resolution spectroscopic measurements. The target we consider is presented in \citet{Gonzalez_2009}. This has a flux of $\sim$100 mJy at mm wavelengths. Due to dust in the source galaxy, it is best to do the observations at such wavelengths. For this purpose, we consider using the Atacama Large Millimetre Array (ALMA). The Bullet Cluster rises to within $\sim$35$^\circ$ of the zenith at this site (minimum airmass $\approx 1.2$).

The parameters we supplied are given in Table \ref{ALMA_inputs}. To clarify the tension with $\Lambda CDM$, it would be necessary to constrain the collision speed to within $\sim$250 km/s (representing an 8\% accuracy if the actual speed is 3000 km/s). This corresponds to determining the redshift difference between the images to 0.1 km/s.

A flux accuracy of 1.5 mJy corresponds to $\sim$2\% accuracy near the peak of the spectral energy distribution. The online calculator suggests that this level of precision can be attained in just over 6 hours under typical weather conditions. Thus, we believe that a night with all 50 of the 12 metre dishes might allow us to constrain, in this case, the proper motion of the main cluster.

%We verified that, for the same input parameters as in the online guide, we got the same results. However, the result is dependent on the browser used to access the Internet, the version of Java and/or the operating system. In fact, one of the authors (IB) presently has two tabs of Internet Explorer giving different exposure time estimates for exactly the same parameters. Thus, the required exposure time may be 2.3 times larger than we obtained (we note that this is very slightly less than $Ln(10)$ and probably indicates an extra/missing factor of 2.3 somewhere in the code, under certain circumstances). Even with this higher estimate for the required exposure time, it should still be possible to detect the MCE.

\begin{table}	 
	\caption{Input parameters used for the ALMA exposure time calculator, available at:
	\newline
	\newline
	$\text{https://almascience.eso.org/proposing/sensitivity-calculator}$
	\newline
	\newline
	The dual polarisation mode should be used as polarisation is unimportant here. The angular resolution does not affect the result, which was 6.17 hours. \newline}
	\label{ALMA_inputs}	
		\begin{tabular}{ll}
			\hline
			Parameter & Value \\
			\hline
			Declination & $-56^\circ$ \\			
			Frequency & 150 GHz \\
			Bandwidth per polarisation & 100 m/s \\
			Water vapour column density & Default: 5$^\text{th}$ octile (1.796 mm) \\ 
			Number of antennae & 50 $\times$ 12 metre \\
			RMS sensitivity & 1.5 mJy \\			
			\hline
		\end{tabular}
\end{table}

In principle, the sub-cluster's motion can also be determined using the MCE. However, we could not find known multiple images with separation close to the East-West direction, the likely direction of the collision. Thus, suitable multiple images would first need to be found around the sub-cluster. This might be accomplished using a fairly deep exposure with the Atacama Pathfinder EXperiment (APEX) or other telescopes. If suitable targets were found, they could be targeted for detailed spectroscopic follow-up. 

The actual direction of motion of each component of the Bullet Cluster is not known for certain. Thus, observers should target multiple images separated in roughly orthogonal directions. Observing more than one object can also minimise systematics associated with details of the source and the lens.

Once suitable targets are found, we believe that a few nights of observations should be sufficient. The field of view might be large enough that multiple images of different sources can be observed in the same pointing, reducing the required telescope time further. Observing the images simultaneously can also reduce systematics associated with changing atmospheric conditions. 

The main difficulty would be in achieving a very accurate calibration of the spectra. However, it is the \emph{relative} redshift between multiple images that is critical for determining the MCE. \emph{Absolute} redshifts are not needed very precisely.

\section{Conclusions}

%The Differential Magnification Effect (DME) described in this work may hamper our ability to determine the tangential velocity of the Bullet Cluster using the Moving Cluster Effect (MCE). However, a judicious choice of source galaxy combined perhaps with a resolved velocity map of it (even at fairly low accuracy) should allow the effect to be greatly reduced and accurately corrected for. The only really problematic double images to handle in this context are caustic images, which we recommend should be avoided (unless some way is found to correct for the DME). 

The Moving Cluster Effect (MCE) may provide an essentially direct method to determine the tangential motion of high-$z$ lensing clusters such as the Bullet Cluster, thereby clarifying the tension that appears to exist with $\Lambda CDM$ \citep{Molnar_2013}. This requires a precise determination of redshift differences between multiple images of the same object.

We expect the MCE to cause multiple images created by the Bullet Cluster to have a redshift velocity difference of $\sim$1 km/s. We find that, for multiple images of a realistic target, this level of accuracy should be feasible with a night on ALMA, using all 50 of its 12 metre dishes. To determine the motions of both the main and the sub-cluster, multiple pointings may be required.

We considered the effect of the time delay between multiple images. In an expanding Universe, this causes them to have different redshifts (the Differential Expansion Effect, DEE). However, this effect is second order in the deflection angle, whereas the MCE is first order. Thus, the DEE can be neglected compared with the MCE.

The Differential Magnification Effect (DME) arises when observing an object with a redshift gradient across it, most likely due to rotation. The precise way in which the magnification varies across the source is different for different images. This leads to them having different mean redshifts. Under plausible circumstances, the effect is large enough that it must be considered when trying to infer the lens motion (Figure \ref{Overall_Effect}).

We consider various methods for determining how the DME affects image redshifts. All techniques require the profiles of spectral lines, if only to estimate the redshift gradient across the image based on the linewidth. If the line profile could be observed in more detail, then one could exploit the fact that the DME and MCE affect the line profile in different ways (Figure \ref{Residuals_k_0_5}). 

Otherwise, the DME could be estimated by determining the parameters which control it (disk scale length \& orientation, maximum line of sight rotation speed \& sense of rotation, how magnification varies with position in the source plane for each image and, to a smaller extent, the source surface density).

The DME is smaller for some sources than for others. We discuss which types of source might reduce the DME in Section \ref{Mitigation}. We believe the best option is to use multiple images of an elliptical galaxy as these are likely to rotate slower than spirals, if at all. In particular, the triple image identified in \citet{Gonzalez_2009} might be a good source to observe.

The DME is larger for lower mass lenses, making it more important for galaxy-galaxy lensing. Measuring peculiar velocities of galaxies using the MCE might thus be very challenging, especially as these are likely smaller than for the Bullet Cluster. 

However, the DME might be easier to observe. This might give more information about the gravitational potential of the lensing galaxy and perhaps the redshift structure of the source. We speculate that the DME might provide a way to estimate the radial gradient in the velocity dispersion of a distant lensed elliptical galaxy.

The authors wish to thank the referee for several very useful comments. Thanks also to Keith Horne and Douglas Buisson for helpful discussions. IB is supported by a Science and Technology Facilities Council studentship.

\newpage

\bibliographystyle{mn2e}
\bibliography{DME_bbl}
\bsp
\end{document}